\documentclass[11pt,a4paper]{article}
\usepackage[utf8]{inputenc}
\usepackage{cite}
\usepackage{hyperref}
\usepackage{subfig}
\usepackage{lipsum}
\usepackage{commath, amssymb, slashed, mathrsfs, bbm, braket}
\usepackage[pdftex,dvipsnames]{xcolor}
\usepackage{xargs}
\usepackage{tikz}
\usepackage{tikz-3dplot}
\usepackage{tikz-cd}
\usepackage{pgfplots}
\usetikzlibrary{shadows,fadings,positioning,calc,arrows,decorations.markings,patterns}

\numberwithin{equation}{section}

\pgfplotsset{compat=1.17}

\begin{document}
\begin{titlepage}
  \thispagestyle{plain}
  \begin{flushright} 
    UUITP -- 52/21
  \end{flushright}
  \vskip 1.5in
  \begin{center}
    {\bf\Large{      
       SYM on Quotients of Spheres and Complex Projective Spaces}}
    \vskip
    0.5in { {Jim Lundin and Lorenzo Ruggeri }
    } \vskip 0.5in {\small{ 
        \emph{Department of Physics and Astronomy, Uppsala University,\\ Box 516, SE-75120 Uppsala, Sweden} 
      }
    }
  \end{center}
  \vskip 0.5in
  \baselineskip 16pt

  \begin{abstract}    
    We introduce a generic procedure to reduce a supersymmetric Yang-Mills (SYM) theory along the Hopf fiber of squashed $S^{2r-1}$ with $U(1)^r$ isometry, down to the $\mathbb{CP}^{r-1}$ base. This amounts to fixing a Killing vector $v$ generating a $U(1)\subset U(1)^r$ rotation and dimensionally reducing either along $v$ or along another direction contained in $U(1)^r$. To perform such reduction we introduce a $\mathbb{Z}_p$ quotient freely acting along one of the two fibers. For fixed $p$ the resulting manifolds $S^{2r-1}/\mathbb{Z}_p\equiv L^{2r-1}(p,\pm 1)$ are a higher dimensional generalization of lens spaces. In the large $p$ limit the fiber shrinks and effectively we find theories living on the base manifold. Starting from $\mathcal{N}=2$ SYM on $S^3$ and $\mathcal{N}=1$ SYM on $S^5$ we compute the perturbative partition functions on $L^{2r-1}(p,\pm 1)$ and, in the large $p$ limit, on $\mathbb{CP}^{r-1}$, respectively for $r=2$ and $r=3$. We show how the reductions along the two inequivalent fibers give rise to two distinct theories on the base. Reducing along $v$ gives an equivariant version of Donaldson-Witten theory while the other choice leads to a supersymmetric theory closely related to Pestun's theory on $S^4$. We use our technique to reproduce known results for $r=2$ and we provide new results for $r=3$. In particular we show how, at large $p$, the sum over fluxes on $\mathbb{CP}^2$ arises from a sum over flat connections on $L^{5}(p,\pm 1)$. Finally, for $r=3$, we also comment on the factorization of perturbative partition functions on non simply connected manifolds.
  \end{abstract}

  \date{}
\end{titlepage}
\setcounter{page}{2}

\tableofcontents

\section{Introduction}
Studying supersymmetric quantum field theories (SQFTs) on curved manifolds led to a vast range of exact results. The work of Pestun \cite{Pestun:2007rz} for $\mathcal{N}=2$ supersymmetric Yang-Mills (SYM) on $S^4$ allowed to explore SQFTs in different dimensions and background geometries (see \cite{Pestun:2016zxk} for a review). In an attempt to collect a class of results into a unique framework, in \cite{Festuccia:2018rew,Festuccia:2019akm} it has been proposed a systematic way of constructing $\mathcal{N}=2$ SYM theories on compact four manifolds with a Killing vector with isolated fixed points. The theories differ by their distribution of either instantons or anti-instantons at the fixed points. As an example Pestun's theory on $S^4$ is obtained placing anti-instantons at one pole and instantons at the other. Considering anti-instantons at both poles gives rise to an equivariant version of Donaldson-Witten (DW) theory \cite{Witten:1988ze}. In our work we consider $\mathcal{N}=2$ SYM on $\mathbb{CP}^2$ and our first goal is to show how both equivariant DW and a supersymmetric theory, closely related to Pestun's\footnote{We will refer to this supersymmetric theory in the rest of the paper as Pestun-like theory. By analogy we will also refer with the same term name to a related theory on $\mathbb{CP}^1$.}, arise from dimensional reducing $\mathcal{N}=1$ SYM on $S^5$. Also for $\mathbb{CP}^2$, DW is obtained placing anti-instantons at all three fixed points, while flipping one of them into an instanton gives a Pestun-like theory. Moreover for both four dimensional theories we are able to compute the perturbative partition functions showing their explicit behaviour at each flux sector. We also study a similar reduction from $S^3$ to $\mathbb{CP}^1$ which we use to test our results. \\

Our starting point in the construction are odd-dimensional spheres. After the work of \cite{Kapustin:2009kz} great progress has been done in recent years in understanding how to compute the partition function of SYM theories on $S^{2r-1}$. For $r=2$, results for an $\mathcal{N}=2$ vector multiplet have been presented for both round and squashed spheres \cite{Jafferis:2010un,Hama:2010av,Hama:2011ea,Imamura:2011wg}. Similar results have been obtained for $r=3$ and an $\mathcal{N}=1$ vector multiplet in \cite{Kallen:2012cs,Kallen:2012va,Imamura:2013xna,Lockhart:2012vp,Kim:2012ava,Kim:2012qf}. The works of~\cite{Pasquetti:2011fj,Beem:2012mb} showed how the partition function for a squashed $S^3$ can be decomposed in two elementary blocks, one for each fixed point of $S^3$. Each factor is given by a copy of the partition function on $\mathbb{C}\times S^1$. The same approach has been conjectured on squashed $S^5$ \cite{Qiu:2014oqa,Tizzano:2014roa} with each factor now written as a partition function on $\mathbb{C}^2\times S^1$. See \cite{Pasquetti:2016dyl} for a review of factorization of more generic manifolds in three and five dimensions. Some results are known also for $r=4$\cite{Polydorou:2017jha,Rocen:2018xwo,Iakovidis:2020znp}.\\

Spheres in odd dimensions can be seen as an Hopf fibration $S^1\hookrightarrow S^{2r-1}\rightarrow\mathbb{CP}^{r-1}$. In this paper we study the dimensional reduction along the Hopf fiber of SYM theories, focusing on $r=2,3$. The first class of theories we consider on $\mathbb{CP}^{r-1}$ are topological twists of $\mathcal{N}=(2,2)$ SYM for $r=2$ and $\mathcal{N}=2$ SYM for $r=3$. In $d=4$ the partition function and other observables compute Donaldson-Witten invariants \cite{Witten:1988ze}. Moreover the $d=2r-2$ manifold admits a $U(1)^{r-1}$ torus action which can be used to define the equivariant version of topological twist. For $r=2,3$ see \cite{Closset:2015rna,Rodriguez-Gomez:2014eza,Bershtein:2015xfa}. We also consider Pestun-like theories on $\mathbb{CP}^{r-1}$. For $r=2$ it has been studied in \cite{Benini:2012ui} and for $r=3$ in \cite{Festuccia:2018rew,Festuccia:2019akm}. Similarly as for the spheres, the partition functions on $\mathbb{CP}^{r-1}$ factorize into elementary blocks defined on $\mathbb{C}^{r-1}$, with each factor coming from one of the $r$ fixed points of the isometry group $U(1)^{r-1}$.\\ 

Having introduced the main objects of interest, in our work we present a systematic way of relating the two, via dimensional reduction along the Hopf fiber:
\begin{itemize}
    \item[(i)]{We take a round $S^{2r-1}$  with $SO(2r)$ isometry group. We fix an arbitrary Killing vector $v$ by choosing a pair of supercharges $\mathbb{Q},\widetilde{\mathbb{Q}}$. The vector $v$ selects a particular $U(1)$ rotation inside the $U(1)^r$ Cartan of the isometry group. The perturbative partition function is a product over $r$ positive integers $n_1,...,n_r$ eigenvalues under each of the $U(1)^r$ along $v$. }
    \item[(ii)]{We consider two different choices for the direction of the Hopf fiber: either the $U(1)$ selected by the Killing vector $v$ or a different combination $\widetilde{U(1)}\subset U(1)^r$. We rewrite the product over $n_1,...,n_r$ as a product over $t,n_2,...,n_r$, with $t$ being the quantum number for either $U(1)$ or $\widetilde{U(1)}$ rotations. We assume these to correspond to $t=\pm n_1+...+n_r$. At fixed $t$, the product over $n_2,...,n_r$ represents two different $(r-1)$-dimensional slices of the same cone in $\mathbb{R}^r$ spanned by positive $(n_1,..,n_r)$.}
    \item[(iii)]{We introduce a quotient of $S^{2r-1}$ by a free $\mathbb{Z}_p$-action, squashed on its $\mathbb{CP}^{r-1}$ base, such that the quotient acts on either of the two Hopf fibers. The resulting manifolds $L^{2r-1}(p,\pm 1)$ are (squashed) higher dimensional analogues of lens spaces. For $L^{3}(p,- 1)$ see \cite{Alday:2012au}\footnote{Notice that in \cite{Alday:2012au} they use a slightly different notation for the $\mathbb{Z}_p$-action and thus the lens space they consider turns out to be $L^3(p,+1)$. This corresponds to $L^3(p,-1)$ in our notations.} while the difference between $L^{3}(p,+1)$ and $L^{3}(p,-1)$ has been studied in \cite{Closset:2017zgf,Closset:2018ghr}. Due to the squashing along the base the Killing vector $v$ generates a rotation which deviates, along the $\mathbb{CP}^{r-1}$ base, from the $U(1)$ of the Hopf fiber. At finite $p$ the SYM partition functions on $L^{2r-1}(p,\pm 1)$ are given by a sum over inequivalent flat connections $\alpha(\mathfrak{m})$. Moreover the quotient by the $\mathbb{Z}_p$-action along the Hopf fibers introduces a projection condition $\pm n_1+n_2+n_3=t=\alpha(\mathfrak{m})\mbox{ mod }p$. At large $p$ the fiber shrinks to a point and the projection condition sets $t=\alpha(\mathfrak{m})$. Because of this, at given $\alpha(\mathfrak{m})$, the perturbative partition functions count a single $(r-1)$-dimensional slice of $n_2,...,n_r$. Due to the different definitions of $t$ in $L^{2r-1}(p,\pm 1)$, we will show how the slices are different in the two reductions. Comparing with the perturbative partition function on $\mathbb{CP}^{r-1}$ we see how, in the limit, the sum over flat connections in $L^{2r-1}(p,\pm 1)$ corresponds to a sum over fluxes in one dimension less. Reducing along the Hopf fiber associated to $v$ we obtain the equivariant version of topologically twisted theories on $\mathbb{CP}^{r-1}$. Instead reducing along $\widetilde{U(1)}$ we find exotic theories on squashed $\mathbb{CP}^{r-1}$, which we match with Pestun-like theories.}
\end{itemize}
We focus on $r=2,3$ where only one\footnote{For $r\geq 4$ more inequivalent choices are allowed.} choice of $\widetilde{U(1)}$ is possible. At each step in the reduction we show how the perturbative partition functions factorize into $r$ pieces\footnote{However we do not study the precise choice of integration contour in the cases we consider.}. In particular we find factorized results on a non simply connected manifold as $L^{2r-1}(p,\pm 1)$ and, at all flux sectors, on $\mathbb{CP}^{r-1}$.\\

For $r=2$, we reproduce an example of such dimensional reductions which appeared in \cite{Benini:2012ui}, where it was shown that the large $p$ limit of the $\mathcal{N}=1$ (round) $L^3(p,-1)$ partition function \cite{Benini:2011nc,Alday:2012au} matches the Pestun-like $\mathcal{N}=(2,2)$ on $\mathbb{CP}^1$. We test the more general procedure reducing using $L^3(p,+1)$ and show how the resulting perturbative partition function matches with the result for $\mathcal{N}=(2,2)$ topologically twisted on $\mathbb{CP}^1$ \cite{Closset:2015rna}. Next\footnote{For $r=3$ a dimensional reduction for a class of toric Sasaki-Einstein manifolds appeared in \cite{Festuccia:2016gul} but missed the contribution of fluxes on the $d=4$ base manifold.} we consider $S^5$ squashed along its $\mathbb{CP}^2$ base and the large $p$ limit of $L^{5}(p,\pm 1)$. The two limits give us results for the perturbative partition functions, at all flux orders, for both $\mathcal{N}=2$ topological twist and an $\mathcal{N}=2$ exotic theory. In both cases the sum over flat connections on $L^{5}(p,\pm 1)$ gives rise to a sum over topological sectors corresponding to fluxes $\alpha(\mathfrak{m})=t$ in the partition function on $\mathbb{CP}^2$. This is different than what was found in \cite{Bershtein:2015xfa,Festuccia:2018rew}, where each flux sector corresponds to several contributions labelled by equivariant fluxes $k^i$, with $i=1,2,3$. Moreover we observe how our results only depend on the Killing vector $v$ on $S^{2r-1}$ and its reduction down to $\mathbb{CP}^{r-1}$.\\

The outline is as follows: in section \secref{sec:two} we introduce the geometry of odd-dimensional squashed spheres. In particular we focus on the two choices of fiber that we will use to reduce with respect to a Killing vector $v$. We also present $L^{2r-1}(p,\pm 1)$ manifolds. We leave to section \secref{sec:three} the relation between $v$ and a choice of supercharges. There we also briefly present the field content of the vector multiplets we consider, and their perturbative partition function on $S^{2r-1}$. These will serve as starting point for sections \secref{sec:cp1} and \secref{sec:cp2} where study, separately, the reductions for $r=2,3$.

\section{Geometry of odd-dimensional spheres}\label{sec:two}
We consider odd-dimensional round spheres $S^{2r-1}$. Most of the results are presented for $r=2$ and $r=3$, however many of the concepts can be extended to generic $r\geq 2$. Considering $S^{2r-1}$ as an Hopf fibration $S^1\hookrightarrow S^{2r-1}\rightarrow\mathbb{CP}^{r-1}$ we study two different choices of fiber with respect to a direction determined by a Killing vector $v$, which is itself determined by a choice of supersymmetry. While the relation between $v$ and supersymmetry will be explained in section \secref{sec:three} here we introduce two sets of coordinates, each adapted to one of the two fibers. Treating the two cases separately, we then introduce a generic squashing which can be set to act either on the $\mathbb{CP}^{r-1}$ base only or on the fiber only. We end the section introducing manifolds obtained as quotients of $S^{2r-1}$ by a free $\mathbb{Z}_p$-action along either of the two fibers. The resulting manifold $S^{2r-1}/\mathbb{Z}_p\equiv L^{2r-1}(p,\pm 1)$ is not simply connected and is a higher dimensional generalization of the $r=2$ lens space. Taking the large $p$ limit of $L^{2r-1}(p,\pm 1)$ we can dimensionally reduce along the two fibers down to $\mathbb{CP}^{r-1}$.

\subsection{Round spheres}
Odd-dimensional spheres $S^{2r-1}$ can be embedded in $\mathbb{C}^r$. Choosing complex coordinates $(z_1,...,z_r)$, the $SO(2r)$-invariant metric can be written as:
\begin{equation}\label{eq:metric.complex}
	ds^2_{S^{2r-1}}=\sum_{i=1}^r|d z_i|^2,\quad\quad \sum_{i=1}^r |z_i|^2=R^2.
\end{equation}
Introducing real coordinates $(\rho_i,\theta_i)$ such that $z_i=\rho_i e^{i\theta_i}$, the metric can be rewritten as:
\begin{equation}\label{eq:metric.real}
	ds^2_{S^{2r-1}}=\sum_{i=1}^r (d\rho_i^2+\rho_i^2d\theta_i^2),\quad\quad \sum_{i=1}^r \rho_i^2=R^2.
\end{equation}
From now on we set $R=1$. Spheres in odd dimensions can be seen as a fibration over $\mathbb{CP}^{r-1}$: $S^1\hookrightarrow S^{2r-1}\rightarrow \mathbb{CP}^{r-1}$. For $r=2$ it corresponds to the Hopf fibration: $S^1\hookrightarrow S^3\rightarrow S^2$. Hence, with a further change of coordinates, their metric can be written as:
\begin{equation}\label{eq:metric.fibration}
	ds^2_{S^{2r-1}}=ds^2_{\mathbb{CP}^{r-1}}+(d\alpha+V)^2,
\end{equation}
where $ds^2_{\mathbb{CP}^{r-1}}$ is the induced Fubini-Study metric, $\alpha$ is the coordinate along the fiber and $V$ is a connection one-form. At this point every choice of fiber is equivalent as they are all related by an $SO(2r)$-rotation. However, as we will explain in detail in the next section, a choice of supersymmetry generators\footnote{Equivalently, as $S^{2r-1}$ are contact manifolds, the choice of supercharges corresponds to a choice of contact structure which uniquely determines a Reeb vector field, which is also Killing.} determines a fixed direction on $S^{2r-1}$. This because the square of the chosen supercharges gives a Killing vector $v$ generating a $U(1)$. To describe this choice we introduce the action of the $U(1)^r\subset SO(2r)$ Cartan of the isometry group on $z_i\rightarrow e^{i\alpha_i}z_i$ for $i=1,...,r$. Denoting $e_i$ the corresponding vector field, the chosen Killing vector field for a round $S^{2r-1}$ is given by:
\begin{equation}\label{eq:killingvec.round}
    v=+e_1+...+e_r.
\end{equation}
With respect to the direction determined by the Killing vector not all choices of fiber are equivalent. In particular we will consider two fibers differing by the action of the first factor of $U(1)$ in the Cartan\footnote{For $r=2$ and $r=3$ all other combinations are related to these two choices, as they can be obtained by shuffling the $z_i$ coordinates and/or flipping all signs. This is not true for $r\geq 4$.}:
\begin{align}\label{eq:fiber.top}
	\mbox{top:}&\quad x^{top}=+e_1+...+e_r,\\
	\label{eq:fiber.ex}
	\mbox{ex:}&\quad x^{ex}=-e_1+...+e_r.
\end{align}
We introduced the notation ``top''and ``ex'' which stands for topologically twisted and exotic, labeling the two different cases associated to the two choices of fiber $x^{top}$ and $x^{ex}$. As we will be interested in dimensionally reducing along a fiber, we will see how topologically twisted theories in $d=2r-2$ are obtained reducing along a direction determined by the Killing vector~\eqref{eq:killingvec.round} differently than exotic theories. This will turn out to be a key point in understanding the two different theories on $\mathbb{CP}^{r-1}$.\\

We now show the explicit changes of coordinates on $S^3$ and $S^5$ adapted to the two fibers~\eqref{eq:fiber.top} and~\eqref{eq:fiber.ex}.

\paragraph{Three sphere:} First we consider $r=2$ and we would like to find different changes of coordinates between~\eqref{eq:metric.real} and~\eqref{eq:metric.fibration}. For this we utilize the choice of basis functions on $S^3$:
\begin{equation}
	\rho_1=\sin\frac{\phi}{2},\quad \rho_2=\cos\frac{\phi}{2},
\end{equation}
along with a choice of angles:
\begin{align}\label{eq:change.asd.3d}
	\mbox{top:}&\quad \theta_1=\frac{1}{2}\left(\alpha-\beta\right),\quad \theta_2=\frac{1}{2}\left(\beta+\alpha\right),\\
	\label{eq:change.flip.3d}
	\mbox{ex:}&\quad \theta_1=\frac{1}{2}\left(\beta-\alpha\right),\quad \theta_2=\frac{1}{2}\left(\beta+\alpha\right).
\end{align}
Inserting the previous relations into~\eqref{eq:metric.real} produces the following metric:
\begin{equation}\label{eq:metric.3d}
	ds^2_{S^3}=\frac{1}{4}\left(d\phi^2+\sin^2\phi d\beta^2+(d\alpha-\cos\phi d\beta)^2\right)=\frac{1}{4}ds^2_{\mathbb{CP}^1}+\frac{1}{4}(d\alpha+V)^2.
\end{equation}
The vector $\frac{\partial}{\partial\alpha}$ generates a rotation along the fiber while we identify the metric of the $\mathbb{CP}^1$ base, along with the one form $V$:
\begin{equation}
	V=-\cos\phi d\beta.
\end{equation}
The two choices of coordinates given in~\eqref{eq:change.asd.3d} and~\eqref{eq:change.flip.3d} both reproduce the same metric~\eqref{eq:metric.3d}. When solving for the $\alpha$ and $\beta$ angles:
\begin{align}
	\mbox{top:}&\quad \beta=-\theta_1+\theta_2,\quad \alpha=+\theta_1+\theta_2,\\
	\mbox{ex:}&\quad \beta=+\theta_1+\theta_2,\quad \alpha=-\theta_1+\theta_2.
\end{align}
we find these two possible inequivalent choices of fiber $\alpha$, differing by a reflection around $\theta_1$, as expected from the two fibers $x^{top}$~\eqref{eq:fiber.top} and $x^{ex}$~\eqref{eq:fiber.ex}.

\paragraph{Five sphere:} Again, considering $r=3$, we need to find two inequivalent changes of coordinates from~\eqref{eq:metric.real} to~\eqref{eq:metric.fibration}. We need to set:
\begin{equation}
	\rho_1=\cos\sigma,\quad \rho_2=\sin\sigma\cos\frac{\phi}{2},\quad\rho_3=\sin\sigma\sin\frac{\phi}{2},
\end{equation}
and:
\begin{align}\label{eq:change.asd}
	\mbox{top:}&\quad\theta_1=+\alpha,\quad\theta_2=\alpha-\frac{1}{2}(\beta+\gamma),\quad\theta_3=\alpha-\frac{1}{2}(\beta-\gamma),\\
	\label{eq:change.flip}
	\mbox{ex:}&\quad\theta_1=-\alpha,\quad\theta_2=\alpha-\frac{1}{2}(\beta+\gamma),\quad\theta_3=\alpha-\frac{1}{2}(\beta-\gamma).
\end{align}
Explicitly substituting into~\eqref{eq:metric.real} gives:
\begin{equation}\begin{split}
	ds^2_{S^5}=&d\sigma^2+\frac{1}{4}\sin^2\sigma(d\phi^2+\sin^2\phi d\gamma^2)+\frac{1}{4}\cos^2\sigma\sin^2\sigma(d\beta+\cos\phi d\gamma)^2+(d\alpha+V)=\\
	=&ds^2_{\mathbb{CP}^2}+(d\alpha+V)^2.
\end{split}\end{equation}
Again $\frac{\partial}{\partial\alpha}$ generates rotation along the fiber. We have defined the one form $V$ as:
\begin{equation}
	V=-\frac{1}{2}\sin^2\sigma(d\beta+\cos\phi d\gamma).
\end{equation}
As for $S^3$ we have found two changes of coordinates which differ by a reflection around $\theta_1$, again in agreement respectively with $x^{top}$~\eqref{eq:fiber.top} and $x^{ex}$~\eqref{eq:fiber.ex}.

\subsection{Squashing}
We are interested in squashed spheres which, in general, break the $SO(2r)$ isometry group of $S^{2r-1}$ to its Cartan $U(1)^r$. The squashed metric in coordinates~\eqref{eq:metric.real} is:
\begin{equation}\label{eq:metric.real.squashed}
	ds^2_{S^{2r-1}}=\sum_{i=1}^r (d\rho_i^2+\rho_i^2d\theta_i^2)+\frac{1}{1-\sum_{i=1}^r a_i^2\rho_i^2}\bigg(\sum_{i=1}^r a_i\rho_i^2 d\theta_i\bigg)^2,\quad\quad\sum_{i=1}^r \rho_i^2=1.
\end{equation}
We define the squashing parameters:
\begin{equation}
    \boldsymbol{\omega}\equiv(\omega_1,...,\omega_r),\quad \omega_i=1+a_i\in\mathbb{R}.
\end{equation}
Setting all $a_i=0$ gives the round sphere.\\

Considering a Hopf fibration, the parameters $a_i$ can be set to make the squashing act only on the base $\mathbb{CP}^{r-1}$, on the fiber, or on a combination of the two. From the definitions of the two chosen fibers $x^{top}$~\eqref{eq:fiber.top} and $x^{ex}$~\eqref{eq:fiber.ex} we know that a squashing acting only on the base is achieved setting:
\begin{align}\label{eq:squashing.base.top}
    \mbox{top:}&\quad+a_1+a_2+...+a_r=0,\\
    \label{eq:squashing.base.ex}
	\mbox{ex:}&\quad-a_1+a_2+...+a_r=0.
\end{align}
To consider a squashing of the fiber instead we require:
\begin{align}
	\mbox{top:}&\quad(+a_1-a_2,...,+a_1-a_r)=(0,...,0),\\
	\mbox{ex:}&\quad(+a_1+a_2,...,+a_1+a_r)=(0,...,0).
\end{align}
We see how a squashing acting only on the fiber imposes strict conditions, relating all squashing parameters $a_i$.
%\begin{align}\label{eq:condition.fiber.asd}	
%	\mbox{top:}&\quad a_1=+a_2=...=+a_r,\\
%	\label{eq:condition.fiber.flip}
%	\mbox{ex:}&\quad a_1=-a_2=...=-a_r.
%\end{align}
In the case of a squashing only on the fiber it is possible to preserve a bigger isometry subgroup, $SU(r)\times U(1)$, where $SU(r)$ is the isometry group of $\mathbb{CP}^{r-1}$ while $U(1)$ parametrizes rotations along either of the two fibers.
%Then recalling the two different changes of coordinates for the topologically twisted~\eqref{eq:change.asd} and the exotic~\eqref{eq:change.flip} fibers (similarly~\eqref{eq:change.flip.3d}  and~\eqref{eq:change.asd.3d} for $S^3$), we find that in both cases the metric~\eqref{eq:metric.fibration} takes the form:
%\begin{equation}
%	ds^2_{S^{2r-1}}=ds^2_{\mathbb{CP}^{r-1}}+\frac{1}{v^2}(d\alpha+V)^2.
%\end{equation}
This case is also particularly interesting as it can be used to dimensionally reduce the manifold along the fiber onto $\mathbb{CP}^{r-1}$, considering the large squashing limit of the fiber. However, as we will dimensionally reduce quotienting by a $\mathbb{Z}_p$-action along a fiber at large $p$, we assume from now on that the squashing parameters are set as in~\eqref{eq:squashing.base.top} and~\eqref{eq:squashing.base.ex} to act only on the $\mathbb{CP}^{r-1}$ base. \\

As we will see in section \secref{sec:three} the perturbative partition function of a vector multiplet\footnote{We will consider $\mathcal{N}=2$ vector multiplets on $S^3$ and $\mathcal{N}=1$ on $S^5$, these will be introduced in section \secref{sec:three}.} on $S^{2r-1}$ factorizes into $r$ factors. Each of these corresponds to a submanifold where the $U(1)^r$ isometry degenerates to a single $U(1)$:
\begin{equation}\begin{split}\label{eq:fixed.fiber}
    &S^3:\quad(\rho_1,\rho_2)=(1,0),\;(0,1),\\
    &S^5:\quad(\rho_1,\rho_2,\rho_3)=(1,0,0),\;(0,1,0),\;(0,0,1).
\end{split}\end{equation}
These $S^1$ fibers are special as they are fixed fibers of a subset $U(1)^{r-1}$ of the full isometry group $U(1)^r$. In a neighbourhood of the fixed fibers the manifold can be identified with a twisted solid torus $\mathbb{C}^{r-1}\times S^1$. We associate inhomogeneous coordinates for the planes $\mathbb{C}^{r-1}$ at each fixed fiber:
\begin{equation}\begin{split}\label{eq:inhomogenous.coordinates}
    &S^3:\quad(1,0),\;(0,1)\rightarrow \left[1,\frac{z_2}{z_1}\right],\;\left[\frac{z_1}{z_2},1\right],\\
    &S^5:\quad(1,0,0),\;(0,1,0),\;(0,0,1)\rightarrow \left[1,\frac{z_2}{z_1},\frac{z_3}{z_1}\right],\;\left[\frac{z_1}{z_2},1,\frac{z_3}{z_2}\right],\;\left[\frac{z_1}{z_3},\frac{z_2}{z_3},1\right].
\end{split}\end{equation}
Focusing on the fixed fiber $(\rho_1,\rho_2,\rho_3)=(1,0,0)$, the twisted identification of the solid torus $\mathbb{C}^2\times S^1$ is: 
\begin{equation}
    \left[1,\frac{z_2}{z_1},\frac{z_3}{z_1}\right]\sim\left[1,\frac{z_2}{z_1}e^{2\pi i\frac{\omega_2}{\omega_1}},\frac{z_3}{z_1}e^{2\pi i\frac{\omega_3 }{\omega_1}}\right],\quad\quad\alpha\sim\alpha+\frac{2\pi}{\omega_1}.
\end{equation}
The description of the other fixed fibers, both on $S^3$ and $S^5$, follows in a similar way.\\

Going to squashed $S^{2r-1}$, the Killing vector field $v$~\eqref{eq:killingvec.round} determined by the choice of supercharges depends on the squashing parameters:
\begin{equation}\label{eq:killingvec.squashed}
    v=\omega_1 e_1 +...+\omega_r e_r.      
\end{equation}
In order to write $v$ in terms of inhomogeneous coordinates, at each fixed fibers, we introduce the vector field $e_t$ corresponding to rotations along the fibers $x^{top}$ and $x^{ex}$: 
\begin{equation}\begin{split}
    \mbox{top:}&\quad e_t^{top}=+e_1+e_2+...+e_r,\\
    \mbox{ex:}&\quad e_t^{ex}=-e_1+e_2+...+e_r.
\end{split}\end{equation}
Considering the fixed fiber $(\rho_1,...,\rho_r)=(1,...,0)$ and substituting for $e_1$, we find:
\begin{equation}\begin{split}\label{eq:killing.fixedfiber}
    \mbox{top:}&\quad v^{top}=+e_t\omega_1+e_2(\omega_2-\omega_1)+...+e_r(\omega_r-\omega_1),\\
    \mbox{ex:}&\quad v^{ex}=-e_t\omega_1+e_2(\omega_2+\omega_1)+...+e_r(\omega_r+\omega_1).
\end{split}\end{equation}
The Killing vectors corresponding to the other fixed fibers follow similarly. 
 
\subsection{Quotients}\label{subsec:modding}
Besides squashing, another action which can be considered on $S^{2r-1}$, squashed on its $\mathbb{CP}^{r-1}$ base, is that of taking the quotient by a freely-acting $\mathbb{Z}_p$ along the fiber:
\begin{equation}\label{eq:Zp.action}
	(z_1,z_2,...,z_r)\rightarrow (z_1 e^{\pm 2\pi i/p},z_2 e^{+2\pi i/p},...,z_r e^{+2\pi i/p}).
\end{equation}
The choice of sign in the first factor corresponds to a quotient acting respectively on the fibers $x^{top}$ and $x^{ex}$. In the case $r=2$ the quotient of $S^3$ by $\mathbb{Z}_p$ is known as (squashed) lens space $L^3(p,\pm 1)$. As we are considering the generalization of such manifolds to higher dimensions we introduce the notation $L^{2r-1}(p,\pm 1)\equiv S^{2r-1}/\mathbb{Z}_p$, where the $\mathbb{Z}_p$-action is that shown in~\eqref{eq:Zp.action} and the $\mathbb{CP}^{r-1}$ base is squashed. Notably these manifolds are not simply connected and thus performing such quotient results into a non-trivial change in the topology of the manifold. In particular on $L^{2r-1}(p,\pm 1)$ the first homotopy group is:
\begin{equation}
	\pi_1(L^{2r-1}(p,\pm 1))\cong \mathbb{Z}_p.
\end{equation}
The free $\mathbb{Z}_p$-action introduces over $L^{2r-1}(p,\pm 1)$ $p$ topologically inequivalent complex line bundles, labeled by flat connections:
\begin{equation}\label{eq:flat.Lens}
    A=\mbox{diag }(A^{m_1}_p,...,A^{m_k}_p).
\end{equation}
Here $0\leq m_i<p$ and the index $i=1,...,k$ counts the Cartan elements of the gauge group $G$. For $r=2$ these have been studied in detail in \cite{Guadagnini:2017lcz}, considering the Heegaard splitting of $L^{3}(p,\pm 1)$ as two solid tori $L^{3}(p,\pm 1)=H_L\cup_f H_R$ identified along their $T^2$ boundary by a homeomorphism $f:\partial H_L\rightarrow\partial H_R$. The flat connections $A_L$ and $A_R$ are related by a gauge transformation $U$ on $\partial H_R$:
\begin{equation}
    f\star A_L=U^{-1}A_R U-iU^{-1}dU.
\end{equation}
Then the flat connection well defined on the entire $L^{2r-1}(p,\pm 1)$ is given by:
\begin{equation}\label{eq:flat.heegaard}
    A=\Bigg\{\begin{matrix} A_L & \mbox{in } H_L\\
    U^{-1}A_R U-iU^{-1}dU & \mbox{in } H_R
    \end{matrix}
\end{equation}
We will show in section \secref{sec:cp1}, following \cite{Guadagnini:2017lcz}, how this affects the evaluation of a Chern-Simons term on $L^{3}(p,\pm 1)$. Moreover, both for $r=2$ and $r=3$, when we will study the partition function for a vector multiplet on $L^3(p,\pm 1)$ and $L^5(p,\pm 1)$, we will have to sum over flat connections differing by their wrapping along the Hopf fiber. To dimensionally reduce onto $\mathbb{CP}^{r-1}$ we can take the large $p$ limit of either $L^{2r-1}(p,+1)$ or $L^{2r-1}(p,-1)$. 

\section{Supersymmetry on $S^{2r-1}$}\label{sec:three}
In the previous section we have presented the geometry of odd-dimensional squashed spheres. We have also shown that on $S^{2r-1}$ for $r=2,3$, a Hopf fibration over $\mathbb{CP}^{r-1}$ can be written choosing two fibers $x^{top}$~\eqref{eq:fiber.top} and $x^{ex}$~\eqref{eq:fiber.ex}, with respect to a direction fixed by a choice of Killing vector $v$~\eqref{eq:killingvec.round}. In this section we motivate how the reductions along either $x^{top}$ or $x^{ex}$ give rise to topologically twisted theories and exotic theories on $\mathbb{CP}^{r-1}$. Hence we show, first, how a specific choice of supersymmetry generators fixes a direction on $S^{2r-1}$ through the Killing vector $v$ given by the square of two supercharges $\mathbb{Q}$, $\widetilde{\mathbb{Q}}$. Second, we explain how the two choices of fiber, in the reduction, affect supersymmetry also on the base manifolds $\mathbb{CP}^{r-1}$. The discussion generalizes to the case of a squashing acting only on the base. We then present the field content of topologically twisted and exotic theories, that is we introduce both $d=3$ $\mathcal{N}=2$ and $d=5$ $\mathcal{N}=1$ vector multiplets using cohomological variables. We conclude this section presenting briefly some known results for the vector multiplets perturbative partition functions $Z_{S^{2r-1}}^{pert}$. These results have been obtained performing a localization computation on $S^3$ \cite{Hama:2010av,Hama:2011ea,Imamura:2011wg} and on $S^5$  \cite{Kallen:2012va,Imamura:2013xna,Lockhart:2012vp,Kim:2012ava,Kim:2012qf}, which showed how, generically, $Z_{S^{2r-1}}^{pert}$ can be expressed as multiple sine function $S_r(i\alpha(\sigma_0)|\boldsymbol{\omega})$. Moreover such functions enjoy a factorization property related directly to the contributions entering the localization computation, with each factor coming from a fixed fiber~\eqref{eq:fixed.fiber}, around which the manifold is locally a twisted solid torus $\mathbb{C}^{r-1}\times S^1$. This brief recap will serve as starting point for the next two sections where we will treat separately $S^3$ and $S^5$.

\subsection{Choice of supercharges}
We will consider $\mathcal{N}=2$ superalgebras on $S^3$ \cite{Jafferis:2010un,Hama:2010av,Hama:2011ea,Imamura:2011wg} and $\mathcal{N}=1$ superalgebras on $S^5$ \cite{Kallen:2012cs,Kallen:2012va,Imamura:2013xna,Lockhart:2012vp,Kim:2012ava}. On the round cases these, and the corresponding bosonic subalgebras, can be determined to be:
\begin{equation}\begin{split}\label{eq:superalgebra}
    &S^3: SU(2)_l\times SU(2|1)\supset SU(2)_l\times SU(2)_r\times U(1)_R,\\
    &S^5 : SU(4|1)\supset SU(4)\times U(1)_R.
\end{split}\end{equation}
The group $SU(2)_l\times SU(2)_r$ and $SU(4)$ are respectively the isometry groups of round $S^3$ and $S^5$, while the subscript $R$ indicates the $R$-symmetry. At the level of Lie algebras we have the isomorphisms:
\begin{equation}
     so(4)\cong su(2)_l\times su(2)_r,\quad  so(6)\cong su(4)
\end{equation}
At this point there is no preferred direction on $S^{2r-1}$ and every choice of $U(1)$ fiber is equivalent. The arbitrary choice we have to do is to select two nilpotent supercharges $Q$ and $\overline{Q}$. If we combine them as:
\begin{equation}
	\mathbb{Q}=Q+\overline{Q},\quad \widetilde{\mathbb{Q}}=Q-\overline{Q},
\end{equation}
the squares $\mathbb{Q}^2$ and $\widetilde{\mathbb{Q}}^2$ give rise to bosonic transformations which include a $U(1)$ rotation. The Killing vector field generating this transformation is chosen to be $v$ defined in~\eqref{eq:killingvec.round}:
\begin{equation}
	v=e_1+...+e_r.
\end{equation}
The subset of the $SO(2r)$ isometry group commuting with the selected $U(1)$ is\footnote{For $S^3$ our choice corresponds to $SU(2)_l$.} $SU(r)$. This is also the isometry group of a $\mathbb{CP}^{r-1}$ base which, however, is not necessarily the base we are reducing onto\footnote{We have been assuming that the reduction is performed introducing a quotient by $\mathbb{Z}_p$ and taking the large $p$ limit. However the following discussion holds also reducing by performing a large squashing acting only on the fiber.}. The two different reductions originate from this observation: we can reduce along a fiber which is either the one specified by $v$ or along a $\widetilde{U(1)}$ contained in the commutant $SU(r)$, as long as the supercharges $\mathbb{Q}$ and $\widetilde{\mathbb{Q}}$ are preserved in the reduction. These two situations correspond respectively to the choices of fiber $x^{top}$~\eqref{eq:fiber.top} and $x^{ex}$~\eqref{eq:fiber.ex}.\\

We now explain the cases of $S^3$ and $S^5$ separately.

\paragraph{Three sphere:}
\begin{itemize}
	\item{Topologically twisted theories: in the round case $x^{top}=v$ and the reduction is along the Killing vector $v$. Notice that $Q$ and $\overline{Q}$ are doublets under $SU(2)_r$ and have $\pm\frac{1}{2}$ charge under the $U(1)$ generated by the Killing vector, as in \cite{Hama:2011ea}. Thus, when reducing, we need to turn on an appropriate $R$-symmetry background field along the fiber. Its reduction has flux in $\mathbb{CP}^1$ such that it cancels the contribution of the spin connection. Before turning on any squashing on the base, the chosen supercharges $\mathbb{Q}$ and $\widetilde{\mathbb{Q}}$ do not generate any isometry on the base manifold $\mathbb{CP}^1$.}
	\item{Exotic theories: reducing along the commutant $\widetilde{U(1)}\subset SU(2)_l$ corresponds to the choice of fiber $x^{ex}$. In this case the supercharges $Q$ and $\overline{Q}$ do not transform under $SU(2)_l$ and there is no $R$-symmetry background field. With respect to this choice of fiber, $\mathbb{Q}$ and $\widetilde{\mathbb{Q}}$ generate transformations on the base $\mathbb{CP}^1$. This is the situation considered in~\cite{Imamura:2011wg} which reduces in $d=2$ to a Pestun-like theory~\cite{Benini:2012ui}.}
\end{itemize}

\paragraph{Five sphere:}
\begin{itemize}
	\item{Topologically twisted theories: again this corresponds to reduce along $x^{top}$ which is equivalent to reduce along the Killing vector $v$. As $Q$ and $\overline{Q}$ have charge $\pm\frac{3}{2}$ we need to turn on a background $R$-symmetry connection. Similarly as for $\mathbb{CP}^1$, the reduced $R$-symmetry field has flux on $\mathbb{CP}^2$ which cancels the spin connection on the base, as in \cite{Bershtein:2015xfa}.}
	\item{Exotic theories: the situation is slightly different for $S^5$ as the full isometry group $SO(6)$ is part of the superalgebra. As before we reduce along the fiber $x^{ex}$ which fixes a $\widetilde{U(1)}$ part of the commutant $SU(3)$. However in this case the supercharges have charge $\pm\frac{1}{2}$ and we need a smaller $R$-symmetry background field. Again, we match this case with Pestun-like theories on $\mathbb{CP}^2$ \cite{Festuccia:2018rew}.}
\end{itemize}

So far we have only considered, for simplicity, the reduction from round $S^{2r-1}$. In sections \secref{sec:cp1} and \secref{sec:cp2} we will be interested in a squashing acting only on the base $\mathbb{CP}^{r-1}$ which breaks the isometry group to its Cartan $U(1)^r$. The condition for the squashing parameters is $\pm a_1+...+a_r=0$ respectively for topologically twisted~\eqref{eq:squashing.base.top} and exotic theories~\eqref{eq:squashing.base.ex}. Notice that while the Killing vector \eqref{eq:killingvec.squashed} is now:
\begin{equation}
	v=\omega_1 e_1+...+\omega_r e_r,
\end{equation}
we can still perform the reduction along the fibers $x^{top}$ and $x^{ex}$. This is possible as the squashing acts only on the base and the fibers $x^{top}$ and $x^{ex}$ are left invariant. Differently than in the round case, $v$ in the reduction along $x^{top}$ generates a $U(1)^{r-1}$ isometry on the base manifolds. This correctly vanishes sending the squashing parameters to zero. 

\subsection{Field content}
Up to now the discussion has been entirely generic with respect to the field content of the theories. We only demanded the existence of two supercharges $\mathbb{Q}$ and $\widetilde{\mathbb{Q}}$ squaring to a particular $U(1)$ direction on $S^{2r-1}$. In order to write down explicitly partition functions in the next sections, we consider $\mathcal{N}=2$ vector multiplet on $S^3$ and an $\mathcal{N}=1$ vector multiplet on $S^5$. The superalgebras are those in \eqref{eq:superalgebra}. Both multiplets consist of a gauge boson $A_\mu$, a real scalar $\sigma$, gauginos $\lambda,\overline{\lambda}$ and an auxiliary scalar $D$. On $S^5$ $\lambda^i$ is a doublet of $SU(2)_R$ while $D_{ij}$ is a triplet. All fields transform in the adjoint of a gauge group $G$. As we want to dimensionally reduce from $S^{2r-1}$ onto the $\mathbb{CP}^{r-1}$ base we would need to consider also the reduction of the vector multiplet fields. However $\mathbb{CP}^2$ is not a spin manifold, due to the non vanishing of the second Stiefel–Whitney class. Hence it is not obvious how to define the reduction of fermions $\lambda,\overline{\lambda}$. Therefore we find it more convenient to introduce cohomological variables which turn all fields into differential forms. The rewriting of fermions gives on $S^3$~\cite{Kallen:2011ny} a zero-form $\alpha$ and a one-form $\Psi$. On $S^5$ instead together with the one-form $\Psi$ we need to include a two-form $\chi$~\cite{Qiu:2016dyj}. All the forms are uncharged under the $R$-symmetry. For the supersymmetry transformations and actions we refer to~\cite{Kallen:2011ny,Qiu:2016dyj} and references therein.

\subsection{Vector multiplet partition function}
The full partition function for a vector multiplet on squashed $S^{2r-1}$ with gauge group $G$ can be written compactly as:
\begin{equation}\label{eq:fullpartition.Sodd}
    Z_{S^{2r-1}}=\int_{\mathfrak{h}} d\sigma_0 e^{-S_{cl}} Z_{S^{2r-1}}^{pert} Z_{S^{2r-1}}^{non-pert}.
\end{equation}
Here $\sigma_0$ is the Coulomb branch parameter integrated over the Cartan subalgebra $\mathfrak{h}$ of the gauge group $G$. Also $Z_{S^{3}}^{non-pert}=1$. Using a localization approach exact results have been computed both for $r=2,3$ for round and squashed $S^{2r-1}$. We consider first the classical part of \eqref{eq:fullpartition.Sodd}. On $S^3$ we turn on a Chern-Simons interaction term \cite{Alday:2012au} while on $S^5$ we use the non $\mathbb{Q}$-exact (nor $\widetilde{\mathbb{Q}}$-exact) SYM action \cite{Qiu:2016dyj}. After evaluating on the BPS locus\footnote{As we will study mainly the perturbative partition functions, on $S^5$ we restrict ourselves to the trivial instanton sector.}, we find:
\begin{align}\label{eq:classical.S3}
    S^3:&\quad e^{-S_{cl}}=e^{i\pi k\varrho_3\mbox{Tr}(\sigma_0^2)},\\
    \label{eq:classical.S5}
    S^5:&\quad e^{-S_{cl}}=e^{-\frac{8\pi^3\varrho_5}{g_{YM}^2}\mbox{Tr}(\sigma_0^2)}.
\end{align}
We have defined 
\begin{equation}
    \varrho_r\equiv\frac{\mbox{Vol}_{S^{2r-1}_{squashed}}}{\mbox{Vol}_{S^{2r-1}}}=\frac{1}{(\omega_1\omega_2...\omega_r)}.
\end{equation}
In this paper we will mainly focus on the perturbative part which can be expressed as multiple sine functions in the following way:
\begin{equation}\label{eq:pert.function}
    Z_{S^{2r-1}}^{pert}=\prod_{\alpha\in roots}S_r(i\alpha(\sigma_0)|\boldsymbol{\omega})
\end{equation}
A possible representation of these multiple sine functions is in terms of multiple gamma functions, turning them into infinite products over the integers $\boldsymbol{n}=(n_1,...,n_r)$:
\begin{equation}\label{eq:multiple.sine}
    S_r(i\alpha(\sigma_0)|\boldsymbol{\omega})=\prod_{n_1,\ldots,n_r=0}^{\infty}\bigg(\boldsymbol{n}\cdot \boldsymbol{\omega} + i\alpha(\sigma_0)\bigg)\prod_{n_1,\ldots,n_r=1}^{\infty}\bigg( \boldsymbol{n}\cdot
    \boldsymbol{\omega} - i\alpha(\sigma_0)\bigg)^{(-1)^{r+1}}
\end{equation}
We see how~\eqref{eq:pert.function} is expressed as an infinite product over $r$ positive integers $n_1,...,n_r$. These represent quantum numbers under the $U(1)^r$ rotations of the modes entering the perturbative partition functions. In particular the $n_i$ count modes of a Fourier expansion in the $\mathbb{C}$-planes into which the background $S^{2r-1}$ can be embedded. For more details on the partition function and multiple sine functions we refer to appendix \ref{app:one}.\\

A different but equivalent way to express this same partition function is to consider the fixed fibers we introduced in section \secref{sec:two}. In the case of $S^{2r-1}$ the manifold is an $S^1$ bundle over $\mathbb{CP}^{r-1}$ which has $r$ fixed points. We can use this fact and express our partition function on $S^{2r-1}$ in terms of $r$ factors, each corresponding to a fixed fiber, giving the following factorized form:
\begin{equation}
    Z^{pert}_{S^{2r-1}}=\prod_{\alpha\in roots}e^{-\mathcal{F}_{eff}[\alpha(\sigma_0),\boldsymbol{\omega}]}\prod_{i=1}^rZ_{\mathbb{C}^{r-1}\times S^1}^{pert}\bigg(\frac{2\pi\alpha(\sigma)}{\omega_i};\frac{2\pi\omega_1}{\omega_i},\ldots,\lor_i,\ldots,\frac{2\pi\omega_r}{\omega_i}\bigg)
\end{equation}
Here we have introduced the perturbative partition function defined on $\mathbb{C}^{r-1}\times S^1$ and we have associated one to each of the fixed fibers. Additionally we have also introduced an effective action $\mathcal{F}_{eff}$ which is not relevant for our work. This factorized fixed point perspective on the partition function of $S^5$, introduced in \cite{Lockhart:2012vp} and further examined in \cite{Kim:2012ava, Nieri:2013vba,Kim:2012qf}, relies on factorization properties of the multiple sine functions shown in \cite{narukawa2004modular}.

\section{Reduction to $\mathbb{CP}^1$}\label{sec:cp1}
In this section we will attempt to lay the groundwork for our approach to lens space reductions on spheres by considering the already well documented example of theories on $S^3$ \cite{Jafferis:2010un,Hama:2010av,Hama:2011ea, Imamura:2011wg} and its lens space $S^3/\mathbb{Z}_p=L^3(p,+1)$ \cite{Benini:2011nc,Benini:2012ui,Alday:2012au}. We will give a slightly more general approach to what is already given in the literature, as we will consider also the possibility of reducing using $L^3(p,-1)$ \cite{Closset:2017zgf,Closset:2018ghr} to get a topologically twisted theory. In section \secref{sec:cp2} we will use this approach to study the case of $S^5$ and two possible lens spaces $S^5/\mathbb{Z}_p=L^5(p,\pm 1)$. To obtain our results we consider $S^3$, squashed along its $\mathbb{CP}^1$ base, and two different fibers $x^{top}$ and $x^{ex}$ with respect to the Killing vector $v$ given by the square of the chosen supercharges $\mathbb{Q},\widetilde{\mathbb{Q}}$. In the unfactorized case, two different expressions for topologically twisted and exotic theories arise when we introduce a rewriting of the perturbative partition function~\eqref{eq:multiple.sine} which counts modes with respect to the $U(1)$ rotations along either $x^{top}$ or $x^{ex}$. The same is true for the factorized case, where, however, we will consider contributions coming from all fixed fibers. Finally we introduce a $\mathbb{Z}_p$ quotient along either of the two fibers. We then need to sum over topological sectors labeled by inequivalent flat connections. In the large $p$ limit these match flux sectors on $\mathbb{CP}^1$.

\subsection{Perturbative partition function on $S^3$}
We consider an $\mathcal{N}=2$ vector multiplet with gauge group $G$ and we start from the perturbative partition function as in~\eqref{eq:multiple.sine}. We are interested in performing some type of dimensional reduction hence it is important to identify the fiber over which we would like to reduce. In the context of the fibers given in \secref{sec:two} and our understanding of the geometry we need to identify what combination of positive integers $n_i$ ``lay'' along a given $S^1$ fiber\footnote{The corresponding $U(1)$ charge of the fiber as generated by the $U(1)$s of the $\mathbb{C}$-planes}. In the case of $S^3$ we have two possible choices:
\begin{align}\label{eq:fiber.3d.asd}
    &t_{top}=+n_1+n_2,\\\label{eq:fiber.3d.flip}
    &t_{ex}=-n_1+n_2,
\end{align}
corresponding to $x^{top}$ and $x^{ex}$. Since we distinguish the $S^1$ fiber and the $S^2$ base we are also free to include a squashing on the base without interfering with the fiber itself. To do so we identify the squashing along the fiber in \eqref{eq:squashing.base.top} and \eqref{eq:squashing.base.ex} and set it to vanish.
In addition we will define the following related equivariant parameters:
\begin{equation}\begin{split}\label{eq:epsilon.3d}
    &\epsilon_{top}=\omega_2-\omega_1,\\
    &\epsilon_{ex}=\omega_2+\omega_1.
\end{split}\end{equation}
As a small comment on the unsquashed limit $\omega_i\rightarrow 1$ as it returns to the round $S^3$, with our interpretation of $\epsilon$ as equivariant parameter we can see that this limit should correspond to the non-equivariant theory. This in turn gives us the non-equivaraiant limit for $\epsilon$ as $\epsilon_{top}\rightarrow 0$ and $\epsilon_{ex}\rightarrow 2$.

\subsubsection{Unfactorized result}
As a starting point we take the result for the total perturbative partition function on $S^3$ from \eqref{eq:multiple.sine}:
\begin{equation}\label{eq:unfactorized.3d}
    Z_{S^{3}}^{pert}=\prod_{\alpha\in roots}\prod_{n_1,n_2\geq 0}\big(n_1\omega_1 + n_2\omega_2+i\alpha(\sigma_0)\big)\prod_{n_1,n_2\geq 1}\big(n_1 \omega_1 + n_2 \omega_2 + i\alpha(\sigma_0)\big)^{-1}
\end{equation}
Notice that this expression is expressed as a product over quantum numbers along the Killing vector $v$~\eqref{eq:killingvec.squashed}. The only difference between the two cases we will consider, is that the conditions for the squashing to be only on the base are different:~\eqref{eq:squashing.base.top} and~\eqref{eq:squashing.base.ex}. To more easily express our partition functions we utilize the following notation:
\begin{equation}
    \Big(\prod_i\Big)^k\Big(\prod_i\Big)^l[f(i)]=\Big(\prod_i f(i)\Big)^{k}\Big(\prod_i f(i)\Big)^l
\end{equation}
We can then take this perturbative determinant and subsitute for $t$ and $\epsilon$ defined in \eqref{eq:fiber.3d.asd},\eqref{eq:fiber.3d.flip} and \eqref{eq:epsilon.3d}:
\begin{equation}\begin{split}\label{eq:unfactorized.3d.asd}
     Z_{S^3}^{pert,top}=&\prod_{\alpha\in roots}\prod_{t\geq n_2\geq 0}\times\bigg(\prod_{t> n_2\geq 1}\bigg)^{-1}\big[\omega_1t+(\omega_2-\omega_1)n_2+i\alpha(\sigma_0)\big]\\
     =&\prod_{\alpha\in roots}\prod_{t\geq n_2\geq 0}\times\bigg(\prod_{t> n_2\geq 1}\bigg)^{-1}\bigg[\epsilon_{top} n_2+i\alpha(\sigma_0)+\Big(1-\frac{\epsilon_{top}}{2}\Big)t\bigg]\\
     =&\prod_{\alpha\in roots}i\alpha(\sigma_0)\prod_{t_{top}= 1}^{\infty}\left[\left(1-\frac{\epsilon_{top}}{2}\right)t_{top}+i\alpha(\sigma_0)\right]\left[\left(1+\frac{\epsilon_{top}}{2}\right)t_{top}+i\alpha(\sigma_0)\right].
\end{split}\end{equation}
\begin{equation}\begin{split}\label{eq:unfactorized.3d.flip}
     Z_{S^3}^{pert,ex}=&\prod_{\alpha\in roots}\prod_{n_2\geq |t|}\times\bigg(\prod_{ n_2+1\geq|t|}\prod_{n_2\geq 1}\bigg)^{-1}\big[-\omega_1t+(\omega_2+\omega_1)n_2+i\alpha(\sigma_0)\big]\\
     =&\prod_{\alpha\in roots}\prod_{n_2\geq |t|}\times\bigg(\prod_{n_2+1\geq|t|}\prod_{n_2\geq 1}\bigg)^{-1}\bigg[\epsilon_{ex} n_2+i\alpha(\sigma_0)-\frac{\epsilon_{ex}}{2}t\bigg]\\
     =&\prod_{\alpha\in roots}\prod_{t_{ex}=-\infty}^{\infty}\left[\frac{\epsilon_{ex}}{2}t_{ex}+i\alpha(\sigma_0)\right].
\end{split}\end{equation}
The first line of both \eqref{eq:unfactorized.3d.asd} and \eqref{eq:unfactorized.3d.asd} shows how this rewriting only depends on the Killing vector $v$ expressed at the fixed fiber $(\rho_1,\rho_2)=(1,0)$ as in~\eqref{eq:killing.fixedfiber}. Another relevant comment is that the quantum number $n_2$, in the two cases, needs to satisfy two different bounds depending on $t$. This will become central when considering each flux sector separately on $\mathbb{CP}^1$.

\subsubsection{Factorized result}
An alternative way to express the $S^3$ partition function is to consider the fixed points of the base $S^2$ manifold and the fiber $S^1$ above each of them. In this way we can find the factorized version of the pertubative partition function:
\begin{equation}
    Z_{S^3}^{pert}=e^{-\mathcal{F}_{eff}}\prod_{l=1}^2Z^{pert}_{\mathbb{C}\times S^1}\bigg(\frac{2\pi \alpha(\sigma_0)}{\omega_l},\frac{2\pi i \omega_k}{\omega_l}\bigg),\quad k=1,2,\quad k\neq l
\end{equation}
Since we know the total perturbative partition function to be a multiple sine function we may utilize its factorization as q-Pochhammer symbols:
\begin{equation}\begin{split}\label{eq:s2.factorization}
    S_2(i\alpha(\sigma_0)|\boldsymbol{\omega})=& e^{\frac{\pi i}{2!}B_{2,2}(i\alpha(\sigma_0)|\boldsymbol{\omega})}\Big(e^{-2\pi\frac{ \alpha(\sigma_0)}{\omega_1}};e^{2\pi\frac{\omega_2}{\omega_1}}\Big)\Big(e^{-2\pi\frac{ \alpha(\sigma_0)}{\omega_2}};e^{2\pi\frac{\omega_1}{\omega_2}}\Big)\\
    =&e^{-\frac{\pi i}{2!}B_{2,2}(i\alpha(\sigma_0)|\boldsymbol{\omega})}\Big(e^{2\pi\frac{ \alpha(\sigma_0)}{\omega_1}};e^{-2\pi\frac{\omega_2}{\omega_1}}\Big)\Big(e^{2\pi\frac{ \alpha(\sigma_0)}{\omega_2}};e^{-2\pi\frac{\omega_1}{\omega_2}}\Big)\\
    =&\left[\Big(e^{-2\pi\frac{ \alpha(\sigma_0)}{\omega_1}};e^{2\pi\frac{\omega_2}{\omega_1}}\Big)\Big(e^{-2\pi\frac{ \alpha(\sigma_0)}{\omega_2}};e^{2\pi\frac{\omega_1}{\omega_2}}\Big)\Big(e^{2\pi\frac{ \alpha(\sigma_0)}{\omega_1}};e^{-2\pi\frac{\omega_2}{\omega_1}}\Big)\Big(e^{2\pi\frac{ \alpha(\sigma_0)}{\omega_2}};e^{-2\pi\frac{\omega_1}{\omega_2}}\Big)\right]^{\frac{1}{2}}
\end{split}\end{equation}
Hence we find three equivalent ways of expressing the factorized partition function. Two of which include $B_{2,2}$ as the effective potential $\mathcal{F}_{eff}$, and a third that consists of square roots of twice as many q-Pochhammer symbols. We use the following definitions for the q-Pochhammer symbols:
\begin{equation}
    \Big(e^{2\pi i \frac{z}{\omega_i}};e^{2\pi i \frac{\omega_l}{\omega_i}}\Big)=\left\{ \begin{array}{ll}
        \prod_{j=0}^{\infty}\Big(1-e^{2\pi i\frac{z}{\omega_i}}e^{2\pi i\frac{\omega_l}{\omega_i}j}\Big) & \mbox{if } \mbox{Im}\big(\frac{\omega_l}{\omega_i}\big)> 0,\vspace{+1em} \\
        \prod_{j=0}^{\infty}\Big(1-e^{2\pi i\frac{z}{\omega_i}}e^{2\pi i\frac{\omega_l}{\omega_i}(-j-1)}\Big)^{-1} & \mbox{if }  \mbox{Im}\big(\frac{\omega_l}{\omega_i}\big)<0.
        \end{array} \right.
\end{equation}
The condition on the ratio of the squashing parameters ensures the expression does not diverge. There is an issue for purely real squashing parameters $\omega_i$. If we attempt to address the round $S^3$ for which the squashings $\omega_i$ are real and equal to $1$, we encounter this very problem. This can be remedied by formally giving small imaginary parts to the squashing parameters $\omega_i$ and then taking the limit as they vanish $\mbox{Im}(\omega_i)\rightarrow 0$ after having performed the expansion using the definition.\\

To present a consistent regularization we start by making an initial assumption, without loss of generality:
\begin{equation}\label{eq:regS3}
    \mbox{Im}\left(\frac{\omega_2}{\omega_1}\right)>0
\end{equation}
If we then assume $\mbox{Re}\left(\omega_1\right)=\mbox{Re}\left(\omega_2\right)=1$, which is the case for the round $S^3$, we can determine the regularization of both q-Pochhammer expressions:
\begin{equation}
    \mbox{Im}\left(\frac{\omega_2}{\omega_1}\right)>0 \quad \Longrightarrow \quad \mbox{Im}\left(\frac{\omega_1}{\omega_2}\right)<0
\end{equation}
Indeed with these assumptions we can also formulate the regularization as:
\begin{equation}
    \mbox{Im}(\omega_2)>\mbox{Im}(\omega_1)
\end{equation}
If we interpret these conditions using the equivariant parameters defined in \eqref{eq:epsilon.3d} we can formulate our regularization as:
\begin{equation}
    \mbox{Im}(\epsilon_{top})>0
\end{equation}
The choice of regularization~\eqref{eq:regS3} tells us to regularize using $\epsilon_{top}$ for both the topologically twisted and exotic fibers\footnote{Assuming $\mbox{Im}(\omega_1/\omega_2)>0$ instead would give $ \mbox{Im}(\epsilon_{top})<0$ for the regularization.}.\\

We wish to deal with the topologically twisted and exotic fibers individually so it is advantageous to express our partition function using their respective epsilons. To do so we can equivalently rewrite the arguments in the q-Pochhammer symbols \cite{Qiu:2013aga} as:
\begin{equation}\begin{split}\label{eq:rotations.3d}
    &\mbox{top:}\quad e^{2\pi i\frac{\omega_2}{\omega_1}}\rightarrow e^{2\pi i\frac{\omega_2-\omega_1}{\omega_1}}=e^{2\pi i\frac{\epsilon_{top}}{\omega_1}}\\
    &\mbox{ex:}\quad e^{2\pi i\frac{\omega_2}{\omega_1}}\rightarrow e^{2\pi i\frac{\omega_2+\omega_1}{\omega_1}}=e^{2\pi i\frac{\epsilon_{ex}}{\omega_1}},
\end{split}\end{equation}
Using the new arguments in \eqref{eq:s2.factorization} we can produce the following factorized expansions for the topologically twisted and exotic fibers:
\begin{equation}\label{eq:factorized.3d.asd}
    \begin{split}
        Z_{S^3}^{pert,top}=&\prod_{\alpha\in roots}\prod_{t=-\infty}^{\infty}\prod_{j=0}^{\infty}\left[\frac{\big(\omega_1 t+\epsilon_{top}j+i\alpha(\sigma_0)\big)\big(\omega_2 t +\epsilon_{top}j+i\alpha(\sigma_0)\big)}{\big(\omega_1 t +\epsilon_{top}(j+1)+i\alpha(\sigma_0)\big)\big(\omega_2 t +\epsilon_{top}(j+1)+i\alpha(\sigma_0)\big)}\right]^{\frac{1}{2}}\\
        =&\prod_{\alpha\in roots}i\alpha(\sigma_0)\prod_{t_{top}=1}^{\infty}\left[\left(1-\frac{\epsilon_{top}}{2}\right) t_{top}+i\alpha(\sigma_0)\right]\left[\left(1+\frac{\epsilon_{top}}{2}\right)t_{top} +i\alpha(\sigma_0)\right]
    \end{split}
\end{equation}
\begin{equation}\label{eq:factorized.3d.flip}
    \begin{split}
        Z_{S^3}^{pert,ex}=&\prod_{\alpha\in roots}\prod_{t=-\infty}^{\infty}\prod_{j=0}^{\infty}\left[\frac{\big(-\omega_1 t+\epsilon_{ex}j+i\alpha(\sigma_0)\big)\big(\omega_2 t -\epsilon_{ex}j+i\alpha(\sigma_0)\big)}{\big(-\omega_1 t +\epsilon_{ex}(j+1)+i\alpha(\sigma_0)\big)\big(\omega_2 t -\epsilon_{ex}(j+1)+i\alpha(\sigma_0)\big)}\right]^{\frac{1}{2}}\\
        =&\prod_{\alpha\in roots}\prod_{t_{ex}=-\infty}^{\infty}\left[\frac{\epsilon_{ex}}{2} t_{ex} +i\alpha(\sigma_0)\right]
    \end{split}
\end{equation}
One can obtain similar expressions before the shifts \eqref{eq:rotations.3d} by substituting for $t=\pm n_1+n_2$ and identifying $j$ with either $n_1$ or $n_2$. Considering the first line in both \eqref{eq:factorized.3d.asd} and \eqref{eq:factorized.3d.flip} we see how in this case the result depends on the Killing vector $v$ computed at both fixed fibers~\eqref{eq:fixed.fiber}. This difference with respect to the unfactorized results will also be true for the $r=3$ case. Of course factorized and unfactorized results match and, after cancellations, we find that this expansion of the perturbative partition function is exactly equivalent to that found in \eqref{eq:unfactorized.3d.asd} and \eqref{eq:unfactorized.3d.flip}. With our expressions reduced to products over the integer $t$ representing Fourier modes along our chosen fibers we are ready to consider the quotient acting on the fibers.

\subsection{Perturbative partition function on $\mathbb{CP}^1$}
With the geometry of lens spaces discussed in section \secref{sec:two} we will use our knowledge to modify the partition function of $S^3$ in order to accommodate the global (topological) effects of a $\mathbb{Z}_p$ quotient. Shown in \eqref{eq:unfactorized.3d}, the partition function on $S^3$ is a product over two positive integers. From the geometry of the lens space we understand that the quotient introduces an identification of segments along a free $S^1$ submanifold. This results in a restriction on the integers that make up the partition function. If we take the quotient to act along the $S^1$ fibers we defined in \eqref{eq:fiber.3d.asd} and \eqref{eq:fiber.3d.flip} we find that it produces the following projection condition for $L^3(p,\pm 1)$:
\begin{equation}\begin{split}\label{eq:projection.3d}
    &t_{top}=+n_1+n_2=\alpha(\mathfrak{m})\mbox{ mod }p,\\
    &t_{ex}=-n_1+n_2=\alpha(\mathfrak{m})\mbox{ mod }p.
\end{split}\end{equation}
Where integers $\alpha(\mathfrak{m})$ that are equivalent modulo $p$ are grouped together as a product, and the equivalence classes $[\mathfrak{m}]$, corresponding to flat connections as in \cite{Alday:2012au}, are then summed over:
\begin{equation}
    Z_{L^3(p,\pm 1)}=\sum_{[\mathfrak{m}]}\int d\sigma_0 e^{-S_{cl}}Z^{pert}_{L^3(p,\pm 1)}(\sigma_0,\mathfrak{m})
\end{equation}
Starting from the classical piece on $L^3(p,\pm 1)$ we need to evaluate the Chern-Simons term on the Abelian flat connections \eqref{eq:flat.heegaard}. Following \cite{Guadagnini:2017lcz} one finds:
\begin{equation}\begin{split}\label{eq:CS.L3}
    S_{cl}[A]&=i\frac{k}{4\pi^2}\int_{L^3(p,\pm 1)}\mbox{Tr}(A\wedge dA)=\\
    &=i\frac{k}{4\pi^2}\int_{H_L}\mbox{Tr}(A_L\wedge dA_L)+i\frac{k}{4\pi^2}\int_{H_R}\mbox{Tr}(A_R\wedge dA_R)+i\frac{k}{4\pi^2}\int_{\partial H_R}\mbox{Tr}(A_R\wedge f\star A_L)=\\
    &=\mp\frac{i\pi k \varrho_3}{p}\mbox{Tr}(\mathfrak{m}^2)
\end{split}\end{equation}
As usual $\mp$ is respectively for two choices of fiber $x^{top}$ and $x^{ex}$ and the Heegaard splitting of $L^3(p,\pm 1)=H_L\cup_f H_R$ has been introduced in section \secref{subsec:modding}. In the last step we have used the fact that flat connections are such that $dA^0_L=dA^0_R=0$. The non vanishing of the Chern-Simons action is due to its failure to be gauge invariant on a manifold with boundaries as $H_R$. To compute the classical contribution for round $L^3(p,\pm 1)$ round, we combine \eqref{eq:CS.L3} with \eqref{eq:classical.S3}:
\begin{equation}\label{eq:classical.L3}
    e^{-S_{cl}}=e^{\frac{i\pi k}{p}\mbox{Tr}(\sigma_0^2\pm\mathfrak{m}^2)}.
\end{equation}
In the limit $p\rightarrow\infty$ we also send $k\rightarrow\infty$ such that the ratio $k/p$ is fixed and reproduce the classical piece for Pestun-like $S^2$ \cite{Benini:2012ui}. We will comment on the reduction along $x^{top}$ of \eqref{eq:classical.L3} after the study of the one-loop determinant.\\

\begin{figure}[!h]
    \centering
    \subfloat[\centering Topologically twisted plot for $t=5$]{{\includegraphics[width=6cm]{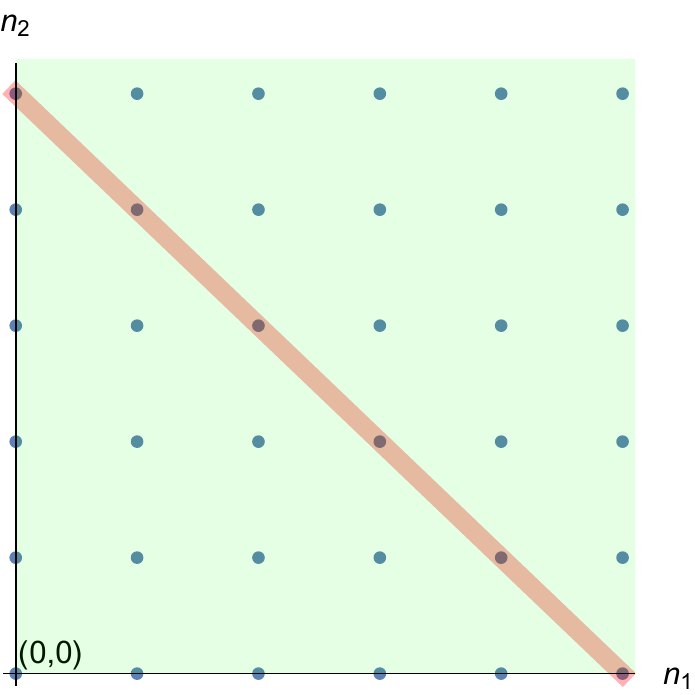} }}%
    \qquad
    \subfloat[\centering Exotic plot for $t=2,-2$]{{\includegraphics[width=6cm]{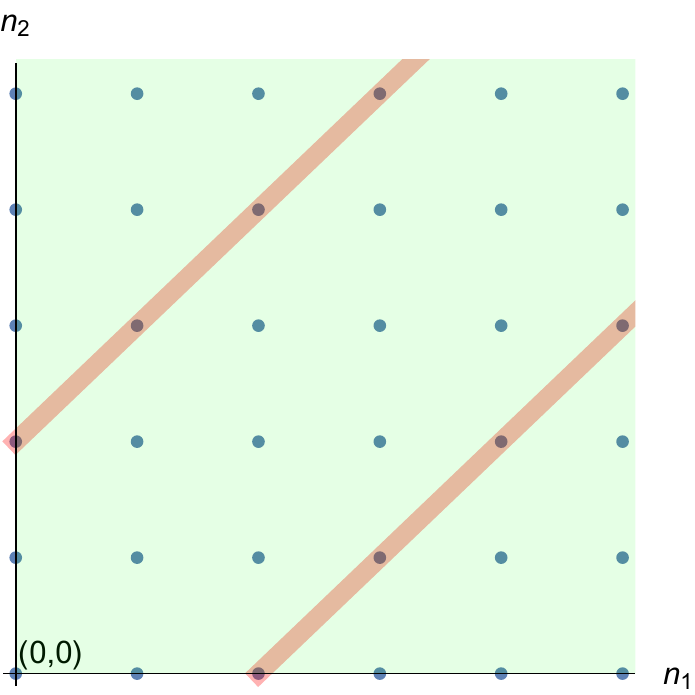} }}%
    \caption{We plot slices at different values of $t=\alpha(\mathfrak{m})$ of the quadrant of $\mathbb{R}^2$ spanned by positive $(n_1,n_2)$ such that $n_1=\mp(n_2-t)$. The orientation of the line is related to the choice of fiber used to reduce.}%
    \label{plot:2dslices}%
\end{figure}

Considering now the perturbative part, we notice how, for large $p$, the modulo operation in \eqref{eq:projection.3d} becomes irrelevant and the statement becomes $n_1+n_2=t_{top}=\alpha(\mathfrak{m})$ and $-n_1+n_2=t_{ex}=\alpha(\mathfrak{m})$. To visualize this projection we can plot a finite part of the points in the $(n_1,n_2)$-plane over which the integers are taken as a product. We show this for different values of $t$ in \autoref{plot:2dslices}. We stress that these need to be considered for the large modding limit $p\rightarrow\infty$, where we will find the result for $\mathbb{CP}^1$.
The diagonal lines in \autoref{plot:2dslices} represent the ``slices'' of values of $n_1,n_2$ such that $\pm n_1+n_2=t=\alpha(\mathfrak{m})$. The effect of this on the unfactorized expression on $S^3$ \eqref{eq:unfactorized.3d.asd} and \eqref{eq:unfactorized.3d.flip} is to restrict $n_2$, at each $t=\alpha(\mathfrak{m})$, to belong to the corresponding slice. However, after simplifications, the results on $S^3$ only depend on $t$\footnote{This will not be the case when considering $r=3$ in the next section. One way to understand this difference is in the appearance of the two products in $S_r(i\alpha(\sigma_0)|\boldsymbol{\omega})$ \eqref{eq:multiple.sine} for $r=2,3$.} and the reduced perturbative partition functions for the two theories on $\mathbb{CP}^1$ are:
\begin{equation}
    Z_{\mathbb{CP}^1}^{pert,top}(\sigma_0,\mathfrak{m})=\Bigg\{\begin{matrix}\prod_{\alpha\in roots}\left[\left(1-\frac{\epsilon_{top}}{2}\right)\alpha(\mathfrak{m})+i\alpha(\sigma_0)\right]\left[\left(1+\frac{\epsilon_{top}}{2}\right)\alpha(\mathfrak{m})+i\alpha(\sigma_0)\right] & \mathfrak{m}>0 \\ \prod_{\alpha\in roots}i\alpha(\sigma_0) & \mathfrak{m}=0\end{matrix}
\end{equation}
\begin{equation}
    Z_{\mathbb{CP}^1}^{pert,ex}(\sigma_0,\mathfrak{m})=\prod_{\alpha\in roots}\left[\frac{\epsilon_{ex}}{2}\alpha(\mathfrak{m})+i\alpha(\sigma_0)\right],\qquad \mathfrak{m}\in \mathbb{Z}^k
\end{equation}
Our result for the topologically twisted theory matches with \cite{Closset:2015rna} up to a constant shift which can be reabsorbed redefining $\sigma_0\rightarrow\sigma_0+i\mathfrak{m}$. Taking into account this redefinition in the classical piece \eqref{eq:classical.L3} we find:
\begin{equation}
     e^{-S_{cl}}=e^{\frac{i\pi k}{p}\mbox{Tr}(\sigma_0^2+2i\sigma_0\mathfrak{m})}.
\end{equation}
As before we take the limit $p\rightarrow\infty$ keeping the ratio $k/p$ fixed. The resulting classical piece in $d=2$ differs from the one in \cite{Closset:2015rna} by a quadratic twisted superpotential \cite{Closset:2014pda}. Regarding $Z_{\mathbb{CP}^1}^{pert,ex}(\sigma_0,\mathfrak{m})$, we can match the result already found in \cite{Benini:2012ui}, up to an overall sign factor dependent on $\alpha(\mathfrak{m})$. As remarked in \cite{Benini:2012ui} this sign dependent factor can possibly be determined by careful examination of the regularization and cancellations of factors in \eqref{eq:factorized.3d.asd} and \eqref{eq:factorized.3d.flip}. \\

Although we have presented both the factorized and unfactorized expressions together there is some extra information when considering the factorized versions. The factorized expressions can be examined factor by factor to find the offset associated to a fixed point.
\begin{equation}\begin{split}
    &l=1\quad\mbox{top: }\left(1-\frac{\epsilon_{top}}{2}\right)\alpha(\mathfrak{m})+i\alpha(\sigma_0)\quad \mbox{ex: }-\frac{\epsilon_{ex}}{2}\alpha(\mathfrak{m})+i\alpha(\sigma_0),\\
    &l=2\quad\mbox{top: }\left(1+\frac{\epsilon_{top}}{2}\right)\alpha(\mathfrak{m})+i\alpha(\sigma_0)\quad \mbox{ex: }+\frac{\epsilon_{ex}}{2}\alpha(\mathfrak{m})+i\alpha(\sigma_0).
\end{split}\end{equation}

\section{Reduction to $\mathbb{CP}^2$}\label{sec:cp2}
In this final section we repeat the computations from the previous section, now for the perturbative partition function of an $\mathcal{N}=1$ vector multiplet on a squashed $S^5$.  So, assuming as before that the squashing acts only on the base, we derive both unfactorized and factorized expressions. We then study the same theory on the five-dimensional lens space $L^5(p,\pm 1)$ whose partition function is expressed as a sum over inequivalent topological sectors corresponding to flat connections. Taking the large $p$ limit we find the reduction onto $\mathbb{CP}^2$, where effectively flat connections correspond to fluxes. Both for topologically twisted and exotic theories, we are able to derive the full perturbative partition function on $\mathbb{CP}^2$, including all flux sectors. Also in this case we provide a factorized and unfactorized expression. It is particularly interesting that our results are presented a sum over a single flux, unlike the approach in~\cite{Bershtein:2015xfa,Festuccia:2018rew}.  

\subsection{Perturbative partition function on $S^5$}
In the previous section we performed the dimensional reduction of $S^3$ by taking the large $p$ limit of $L^3(p,\pm 1)$. In this section we consider the similar procedure on $S^5$ and reduce onto $\mathbb{CP}^2$ by taking a large $\mathbb{Z}_p$ quotient along the fibers corresponding to the topologically twisted~\eqref{eq:fiber.top} and exotic~\eqref{eq:fiber.ex} cases. On top of the quotient we introduce a squashing acting only on the $\mathbb{CP}^2$ base. The three squashing parameters $a_i$ need to satisfy $\pm a_1+a_2+a_3=0$ where the $\pm$ is, as usual, respectively for the topologically twisted or exotic fibers. We can also relate these to equivariant parameters $\epsilon_1,\epsilon_2$. The definitions differ for the two cases:
\begin{equation}\begin{split}\label{eq:epsilon5d}
    \mbox{top:}&\quad\epsilon_1^{top}=\omega_2-\omega_1,\quad\epsilon_2^{top}=\omega_3-\omega_1,\\
    \mbox{ex:}&\quad\epsilon_1^{ex}=\omega_2+\omega_1,\quad\epsilon_2^{ex}=\omega_3+\omega_1.
\end{split}\end{equation}
We notice that the unsquashed limit corresponds to $\epsilon_1^{top}=\epsilon_2^{top}=0$ and $\epsilon_1^{ex}=\epsilon_2^{ex}=2$. In the following we will call in both cases the equivariant parameters simply $\epsilon_1,\epsilon_2$, however they need to be intended as defined above. 

\subsubsection{Unfactorized result}
We have reviewed how to write the full perturbative one-loop determinant on $S^5$ for an $\mathcal{N}=1$ vector multiplet with gauge group $G$ as a triple sine function:
\begin{equation}
    Z_{S^5}^{pert}=\prod_{\alpha\in roots}\prod_{n_1,n_2,n_3\geq 0}\bigg(n_1\omega_1+n_2\omega_2+n_3\omega_3+i\alpha(\sigma_0)\bigg)\prod_{n_1,n_2,n_3\geq 1}\bigg(n_1\omega_1+n_2\omega_2+n_3\omega_3+i\alpha(\sigma_0)\bigg)
\end{equation}
This expression is valid both for topologically twisted and exotic reductions. We also notice that the product is taken over quantum numbers which are eigenvalues along the Killing vector $v$~\eqref{eq:killingvec.squashed}. A first rewriting\footnote{For notational purposes, in the following expression for the perturbative partition function, we will keep the product over roots implicit.} can be obtained by introducing, similarly as for $S^3$, the quantum number for rotations respectively along the fiber $x^{top}$ and $x^{ex}$:
\begin{align}
    &t_{top}=+n_1+n_2+n_3,\\
    &t_{ex}=-n_1+n_2+n_3.
\end{align}
In the rest we will denote both $t_{top}$ and $t_{ex}$ as $t$. Then, in both cases, we can choose to rewrite $Z_{S^5}^{pert}$ in terms of $t$ and two out of the three quantum numbers $(n_1,n_2,n_3)$. Choosing to substitute for $n_1$, we find for the topologically twisted and exotic cases:
\begin{equation}\begin{split}\label{eq:unfactorized.5d.asd}
     Z_{S^5}^{pert,top}=&\prod_{t\geq n_2+n_3}\prod_{n_2,n_3\geq 0}\times\prod_{t\geq n_2+n_3+1}\prod_{n_2,n_3\geq 1}\bigg(\omega_1t+(\omega_2-\omega_1)n_2+(\omega_3-\omega_1)n_3+i\alpha(\sigma_0)\bigg)\\
     =&\prod_{t\geq n_2+n_3}\prod_{n_2,n_3\geq 0}\times\prod_{t\geq n_2+n_3+1}\prod_{n_2,n_3\geq 1}\bigg(\epsilon_1 n_2+\epsilon_2 n_3+i\alpha(\sigma_0)+\bigg(1-\frac{\epsilon_1+\epsilon_2}{3}\bigg)t\bigg).
\end{split}\end{equation}
\begin{equation}\begin{split}\label{eq:unfactorized.5d.flip}
     Z_{S^5}^{pert,ex}=&\prod_{t\leq n_2+n_3}\prod_{n_2,n_3\geq 0}\times\prod_{t\leq n_2+n_3+1}\prod_{n_2,n_3\geq 1}\bigg(-\omega_1t+(\omega_2+\omega_1)n_2+(\omega_3+\omega_1)n_3+i\alpha(\sigma_0)\bigg)\\
     =&\prod_{t\leq n_2+n_3}\prod_{n_2,n_3\geq 0}\times\prod_{t\leq n_2+n_3+1}\prod_{n_2,n_3\geq 1}\bigg(\epsilon_1 n_2+\epsilon_2 n_3+i\alpha(\sigma_0)+\bigg(1-\frac{\epsilon_1+\epsilon_2}{3}\bigg)t\bigg).
\end{split}\end{equation}
Similar re-writings can be obtained substituting for $n_2$ and $n_3$. What is entering the expressions above is the Killing vector written in inhomogenous coordinates at the fixed fiber $(\rho_1,\rho_2,\rho_3)=(1,0,0)$, as in~\eqref{eq:killing.fixedfiber}. We call such expressions unfactorized as they capture ``globally'' all the modes counted by the one-loop determinant on the squashed $S^5$. Also, as for $S^3$, we find that the quantum numbers $n_2,n_3$ need to satisfy, in the two cases, two different bounds depending on $t$. These bounds will be those determining which modes are counted at each flux sector on $\mathbb{CP}^2$.

\subsubsection{Factorized result}
Another approach is to compute the one-loop determinant by summing local contributions around each fixed fiber~\eqref{eq:fixed.fiber} of $S^5$ where, locally, the manifold is equivalent to a twisted solid torus $\mathbb{C}^2\times S^1$. At each fixed fiber the $U(1)^3$ isometry group of the squashed $S^5$ degenerates to a single $U(1)$ whose action is determined by the choice of inhomogenous coordinates~\eqref{eq:inhomogenous.coordinates}. Thus the full perturbative part schematically is given by:
\begin{equation}
    Z_{S^5}^{pert}=e^{-\mathcal{F}_{eff}}\prod_{l=1}^3 Z^{pert,l}_{\mathbb{C}^2\times S^1}\bigg(\frac{2\pi \alpha(\sigma_0)}{\omega_l},\frac{2\pi i\omega_k}{\omega_l},\frac{2\pi i\omega_m }{\omega_l}\bigg),\quad k,m=1,2,3,\quad k,m\neq l\quad m \neq k.
\end{equation}
Here $\mathcal{F}_{eff}$ is an effective prepotential. The factorization property \cite{narukawa2004modular} of the multiple sine functions $S_r$ is well known \eqref{eq:Sr.factorization} and for $r=3$ we find an expression in terms of q-Pochhammer symbols \eqref{eq:q.Pochhammer}:
\begin{equation}\begin{split}\label{eq:S3.Pochhammer}
   S_3(i\alpha(\sigma_0)|\boldsymbol{\omega})&=e^{-\frac{\pi i}{6}B_{3,3}(i\alpha(\sigma_0)|\boldsymbol{\omega})}\bigg[(e^{-2\pi  \frac{\alpha(\sigma_0)}{\omega_1}};e^{2\pi i \frac{\omega_2}{\omega_1}},e^{2\pi i \frac{\omega_3}{\omega_1}})\times(\mbox{2 cyclic permutations on $\omega_i$})\bigg]\\
   &=e^{+\frac{\pi i}{6}B_{3,3}(i\alpha(\sigma_0)|\boldsymbol{\omega})}\bigg[(e^{+2\pi  \frac{\alpha(\sigma_0)}{\omega_1}};e^{-2\pi i \frac{\omega_2}{\omega_1}},e^{-2\pi i \frac{\omega_3}{\omega_1}})\times(\mbox{2 cyclic permutations on $\omega_i$})\bigg]\\
   &=(e^{-2\pi\frac{\alpha(\sigma_0)}{\omega_1}};e^{2\pi i \frac{\omega_2}{\omega_1}},e^{2\pi i \frac{\omega_3}{\omega_1}})^{\frac{1}{2}}(e^{-2\pi\frac{\alpha(\sigma_0)}{\omega_1}};e^{-2\pi i \frac{\omega_2}{\omega_1}},e^{-2\pi i \frac{\omega_3}{\omega_1}})^{\frac{1}{2}}\times(\mbox{2 cyclic permutations on $\omega_i$})
\end{split}\end{equation}
Naively such expressions would seem to depend only on local data however things are more subtle. These infinite products need to be regularized, as explained in~\eqref{eq:q.Pochhammer}, and that is when one is required to patch the local information consistently. For $r=3$ we define four different regularizations:
\begin{equation}\label{eq:F9.Fluder}
 (e^{2\pi i\frac{z}{\omega_i}};e^{2\pi i\frac{\omega_l}{\omega_i}},e^{2\pi i\frac{\omega_m}{\omega_i}}) = \left\{ \begin{array}{ll}
        \prod_{j,k=0}^{\infty}(1-e^{2\pi i\frac{z}{\omega_i}}e^{2\pi i\frac{\omega_l}{\omega_i}j}e^{2\pi i\frac{\omega_m}{\omega_i}k}) & \mbox{if } \mbox{Im}\big(\frac{\omega_l}{\omega_i}\big),\mbox{Im}\big(\frac{\omega_m}{\omega_i}\big)> 0,\vspace{+1em} \\
        \prod_{j,k=0}^{\infty}(1-e^{2\pi i\frac{z}{\omega_i}}e^{2\pi i\frac{\omega_l}{\omega_i}(-j-1)}e^{2\pi i\frac{\omega_m}{\omega_i}k})^{-1} & \mbox{if } \mbox{Im}\big(\frac{\omega_m}{\omega_i}\big)> 0>\mbox{Im}\big(\frac{\omega_l}{\omega_i}\big),\vspace{+1em}\\
       \prod_{j,k=0}^{\infty}(1-e^{2\pi i\frac{z}{\omega_i}}e^{2\pi i\frac{\omega_l}{\omega_i}j}e^{2\pi i\frac{\omega_m}{\omega_i}(-k-1)})^{-1} & \mbox{if } \mbox{Im}\big(\frac{\omega_l}{\omega_i}\big)>0> \mbox{Im}\big(\frac{\omega_m}{\omega_i}\big),\vspace{+1em}\\
        \prod_{j,k=0}^{\infty}(1-e^{2\pi i\frac{z}{\omega_i}}e^{2\pi i\frac{\omega_l}{\omega_i}(-j-1)}e^{2\pi i\frac{\omega_m}{\omega_i}(-k-1)}) & \mbox{if } 0>\mbox{Im}\big(\frac{\omega_l}{\omega_i}\big),\mbox{Im}\big(\frac{\omega_m}{\omega_i}\big).
        \end{array} \right.
\end{equation}
Following~\cite{Chang:2017cdx} we can assume without loss of generality:
\begin{equation}\label{eq:regularization.omegas}
    \mbox{Im }\left(\frac{\omega_2}{\omega_1}\right),\mbox{ Im}\left(\frac{\omega_3}{\omega_1}\right),\mbox{ Im}\left(\frac{\omega_2}{\omega_3}\right)>0.
\end{equation}
Focusing on the round $S^5$ we can simplify the above expressions setting $\mbox{Re}(\omega_1)=\mbox{Re}(\omega_2)=\mbox{Re}(\omega_3)=1$ and considering a small imaginary part:
\begin{equation}
    \mbox{Im}(\omega_i)=a_i.
\end{equation}
Then in this case the regularization in~\eqref{eq:regularization.omegas} reduces to:
\begin{equation}\begin{split}\label{eq:regularization.fp.omegas}
    &l=1: \mbox{ Im}(\omega_2/\omega_1)=a_2-a_1>0,\quad\quad  \mbox{Im}(\omega_3/\omega_1)=a_3-a_1>0,\\
    &l=2:\mbox{ Im}(\omega_1/\omega_2)=a_1-a_2<0,\quad\quad \mbox{Im}(\omega_3/\omega_2)=a_3-a_2<0,\\
    &l=3:\mbox{ Im}(\omega_2/\omega_3)=a_2-a_3>0,\quad\quad\mbox{Im}(\omega_1/\omega_3)=a_1-a_3<0.
\end{split}\end{equation}
On round $S^5$ the equivariant parameters defined for the topologically twisted case $\epsilon_1^{top}=\omega_2-\omega_1=a_2-a_1$ and $\epsilon_2^{top}=\omega_3-\omega_1=a_3-a_1$, are purely imaginary and can be used to determine completely the regularizations. The choice in~\eqref{eq:regularization.omegas} corresponds to set\footnote{Notice that the relation in terms of $\epsilon_{1,2}^{ex}$ would be different using~\eqref{eq:regularization.omegas}. Instead one can use $(\overline{z}_1,z_2,z_3)$ coordinates, assuming $\mbox{Im }(\omega_2/\overline{\omega}_1),\mbox{ Im}(\omega_3/\overline{\omega}_1),\mbox{ Im}(\omega_2/\omega_3)>0$. Then the regularization would depend on $\mbox{Im}(\epsilon^{ex}_{1,2})$.}:
\begin{equation}\label{eq:regularization.epsilons}
    \mbox{Im }(\epsilon_1^{top})>\mbox{ Im }(\epsilon_2^{top})>0.
\end{equation}
Then in terms of the signs of the imaginary parts of $\epsilon^{top}_{1,2}$ the choice of regularization~\eqref{eq:regularization.omegas} becomes:
\begin{equation}\begin{split}\label{eq:regularization.fp.epsilons}
    &l=1: \mbox{ Im}(\epsilon_1^{top})>0,\quad\quad\quad\quad\mbox{ Im}(\epsilon_2^{top})>0\quad\quad\quad\quad\longrightarrow\quad ++,\\
    &l=2:\mbox{ Im}(-\epsilon_1^{top})<0,\;\quad\quad\quad \mbox{ Im}(\epsilon_2^{top}-\epsilon_1^{top})<0\quad\longrightarrow\quad --,\\
    &l=3:\mbox{ Im}(\epsilon_1^{top}-\epsilon_2^{top})>0,\;\quad\mbox{ Im}(-\epsilon_2^{top})<0\quad\quad\quad\longrightarrow \quad +-,
\end{split}\end{equation}
where the definition of $+/-$ regularization is simply a shortening for the four expressions in~\eqref{eq:F9.Fluder}. A small squashing does not affect the regularization and so the chosen distribution of $+/-$ regularizations is a valid assumption also for a squashed sphere, picking either $x^{top}$ as fiber or $x^{ex}$. This assumption is consistent with what is done in $d=5$ for the topologically twisted case in~\cite{Kim:2012qf} and, as we will show below, with the reduction to $d=4$ of the exotic theory \cite{Festuccia:2018rew}. We notice that the parameters determining the regularization correspond to the imaginary part of the Killing vector field $v^{top}$ written in terms of inohomogenous coordinates\footnote{Again, using $(\overline{z}_1,z_2,z_3)$ coordinates makes $v^{ex}$ the relevant Killing vector for the regularization.} at each fixed fiber~\eqref{eq:killing.fixedfiber}.\\

Having in mind the reduction to $\mathbb{CP}^2$ we find it convenient to express the arguments of the q-Pochhammer symbols in~\eqref{eq:S3.Pochhammer} in terms of $\epsilon_i$. For $l=1$ we can equivalently rewrite the arguments of the q-Pochhammer \cite{Qiu:2013aga} for the two choices of fiber as:
\begin{equation}\begin{split}\label{eq:rotations}
    &\mbox{top:}\quad \bigg(e^{2\pi i\frac{\omega_2}{\omega_1}},e^{2\pi i\frac{\omega_3}{\omega_1}}\bigg)\rightarrow \bigg(e^{2\pi i\frac{\omega_2-\omega_1}{\omega_1}},e^{2\pi i\frac{\omega_3-\omega_1}{\omega_1}}\bigg)=\bigg(e^{2\pi i\frac{\epsilon_1}{\omega_1}},e^{2\pi i\frac{\epsilon_2}{\omega_1}}\bigg),\\
    &\mbox{ex:}\quad \bigg(e^{2\pi i\frac{\omega_2}{\omega_1}},e^{2\pi i\frac{\omega_3}{\omega_1}}\bigg)\rightarrow \bigg(e^{2\pi i\frac{\omega_2+\omega_1}{\omega_1}},e^{2\pi i\frac{\omega_3+\omega_1}{\omega_1}}\bigg)=\bigg(e^{2\pi i\frac{\epsilon_1}{\omega_1}},e^{2\pi i\frac{\epsilon_2}{\omega_1}}\bigg),
\end{split}\end{equation}
recalling that $\epsilon_i$ are defined differently in the two cases \eqref{eq:epsilon5d}. Similar rewritings can be found for $l=2,3$. Introducing local equivariant deformations parameters $\epsilon'_i$ we find for the numerators on the right hand sides of~\eqref{eq:rotations}:
\begin{center}
  \begin{tabular}{ c | c | c | c }
    %\hline
    \mbox{top} & $l=1$ & $l=2$ & $l=3$  \\ \hline
    $\epsilon'_1$ & $\epsilon_1$ & $-\epsilon_1$ & $\epsilon_1-\epsilon_2$   \\ \hline
    $\epsilon'_2$ &  $\epsilon_2 $ & $ \epsilon_2-\epsilon_1$ & $ -\epsilon_2 $ 
    %\hline
  \end{tabular}
\end{center}
\begin{center}
  \begin{tabular}{ c | c | c | c }
    %\hline
    \mbox{ex} & $l=1$ & $l=2$ & $l=3$  \\ \hline
    $\epsilon'_1$ & $\epsilon_1$ & $\epsilon_1$ & $\epsilon_1-\epsilon_2$   \\ \hline
    $\epsilon'_2$ &  $\epsilon_2 $ & $\epsilon_2-\epsilon_1$ & $\epsilon_2$ 
    %\hline
  \end{tabular}
\end{center}
Using in both cases the same regularization, as for the round $S^5$, we can rewrite~\eqref{eq:S3.Pochhammer} as:
\begin{equation}\begin{split}\label{eq:2.62asd}
    Z_{S^5}^{pert,top}&=\prod_{t=-\infty}^{\infty}\prod_{i,j=0}^{\infty}\left(\omega_1 t+\epsilon_1 i +\epsilon_2 j+i\alpha(\sigma_0))\right)^{\frac{1}{2}}\left(\omega_1 t+\epsilon_1 (i+1) +\epsilon_2 (j+1)+i\alpha(\sigma_0)\right)^{\frac{1}{2}}\\
    &\left(\omega_2 t-\epsilon_1(-i-1)+(\epsilon_2-\epsilon_1) (-j-1)+i\alpha(\sigma_0)\right)^{\frac{1}{2}}\left(\omega_2 t-\epsilon_1 (-i) +(\epsilon_2-\epsilon_1)(-j)+i\alpha(\sigma_0)\right)^{\frac{1}{2}}\\
    &\left(\omega_3 t+(\epsilon_1-\epsilon_2)i-\epsilon_2 (-j-1)+i\alpha(\sigma_0)\right)^{-\frac{1}{2}}\left(\omega_3 t+(\epsilon_1-\epsilon_2)(i+1) -\epsilon_2(-j)+i\alpha(\sigma_0)\right)^{-\frac{1}{2}}.
\end{split}\end{equation}
\begin{equation}\begin{split}\label{eq:2.62flip}
    Z_{S^5}^{pert,ex}&=\prod_{t=-\infty}^{\infty}\prod_{i,j=0}^{\infty}\left(-\omega_1 t+\epsilon_1 i +\epsilon_2 j+i\alpha(\sigma_0))\right)^{\frac{1}{2}}\left(-\omega_1 t+\epsilon_1 (i+1) +\epsilon_2 (j+1)+i\alpha(\sigma_0)\right)^{\frac{1}{2}}\\
    &\left(\omega_2 t+\epsilon_1(-i-1)+(\epsilon_2-\epsilon_1) (-j-1)+i\alpha(\sigma_0)\right)^{\frac{1}{2}}\left(\omega_2 t+\epsilon_1 (-i)+(\epsilon_2-\epsilon_1)(-j)+i\alpha(\sigma_0)\right)^{\frac{1}{2}}\\
    &\left(\omega_3 t+(\epsilon_1-\epsilon_2)i+\epsilon_2 (-j-1)+i\alpha(\sigma_0)\right)^{-\frac{1}{2}}\left(\omega_3 t+(\epsilon_1-\epsilon_2)(i+1) +\epsilon_2(-j)+i\alpha(\sigma_0)\right)^{-\frac{1}{2}}.
\end{split}\end{equation}
As for $S^3$ similar expressions before the shifts \eqref{eq:rotations} can be found substituting for  $t$ and identifying $i,j$ with two of $n_1,n_2,n_3$. The result in the topologically twisted case matches with~\cite{Kim:2012qf}. We point out that each pair of factors depends on the Killing vectors $v^{top}$ and $v^{ex}$ written in inhomogenous coordinates at each fiber. Below we will show how dimensionally reducing these expressions we find the factorized perturbative partition function on $\mathbb{CP}^2$.

\subsection{Perturbative partition function on $\mathbb{CP}^2$}
As we anticipated previously, to dimensionally reduce onto $\mathbb{CP}^2$, we can perform a $\mathbb{Z}_p$ quotient acting freely on the fiber and take the large $p$ limit. At finite $p$ the partition function computed on the manifold $L^5(p,\pm 1)$ localizes to a set of inequivalent flat connections \eqref{eq:flat.Lens}, specified by holonomies:
\begin{equation}
    A=\mbox{diag }(A^{m_1}_p,...,A^{m_k}_p),
\end{equation}
Where $0\leq m_i<p$ with $i=1,...,k$ and $k$ is the rank of the gauge group $G$. At each topological sector we need to integrate over the covariantly constant scalar $\sigma_0$. The partition function is then a sum over $\mathfrak{m}=\mbox{diag }(m_1,...,m_k)$:
\begin{equation}
    Z_{L^5(p,\pm 1)}=\sum_{[\mathfrak{m}]}\int d\sigma_0 e^{-S_{cl}}Z^{pert}_{L^5(p,\pm 1)}(\sigma_0,\mathfrak{m})Z^{non-pert}_{L^5(p,\pm 1)}(\sigma_0,\mathfrak{m})
\end{equation}
The projection condition for modes on $L^5(p,\pm 1)$ is:
\begin{equation}\begin{split}\label{eq:projection.5d}
    \mbox{top: }\quad &t_{top}=+n_1+n_2+n_3=\alpha(\mathfrak{m})\mbox{ mod }p,\\
    \mbox{ex: }\quad &t_{ex}=-n_1+n_2+n_3=\alpha(\mathfrak{m})\mbox{ mod }p.
\end{split}\end{equation}
At finite $p>1$ and given flux sector $\mathfrak{m}$ the unfactorized perturbative partition function is obtained from~\eqref{eq:unfactorized.5d.asd} and~\eqref{eq:unfactorized.5d.flip} simply by changing the range of $t$ as in~\eqref{eq:projection.5d}. Similar expressions can be obtained for the factorized form. With this result we are able to show how, on a non-simply connected manifold as $L^5(p,\pm 1)$, the perturbative partition function factorizes for each flat connection. This improves the result about factorization in $d=5$ of \cite{Qiu:2014oqa,Pasquetti:2016dyl}. The classical part \eqref{eq:classical.S5} on round $L^5(p,\pm 1)$ at the trivial instanton sector becomes:
\begin{equation}
    e^{-S_{cl}}=e^{-\frac{8\pi^3}{pg_{YM}^2}\mbox{Tr}(\sigma_0^2+f(\sigma,\mathfrak{m}))}.
\end{equation}
The function $f(\sigma,\mathfrak{m})$ can be determined generalizing the approach of \cite{Guadagnini:2017lcz} to higher dimensional lens spaces and we leave it for a future study. In the large $p$ limit, keeping $pg^2_{YM}$ constant\footnote{Notice that, in five dimensions, the Yang-Mills coupling $g^2_{YM}$ has the dimension of a length and it can be related to the radius of the $S^1$ which can be added to $S^5$ to give a six dimensional theory on $S^5\times S^1$ \cite{Kallen:2012va}}, the classical piece reduces to that computed on $\mathbb{CP}^2$. \\

\begin{figure}[h!]
    \centering
    \subfloat[\centering Topologically twisted plot for $t=5$]{{\includegraphics[width=6cm]{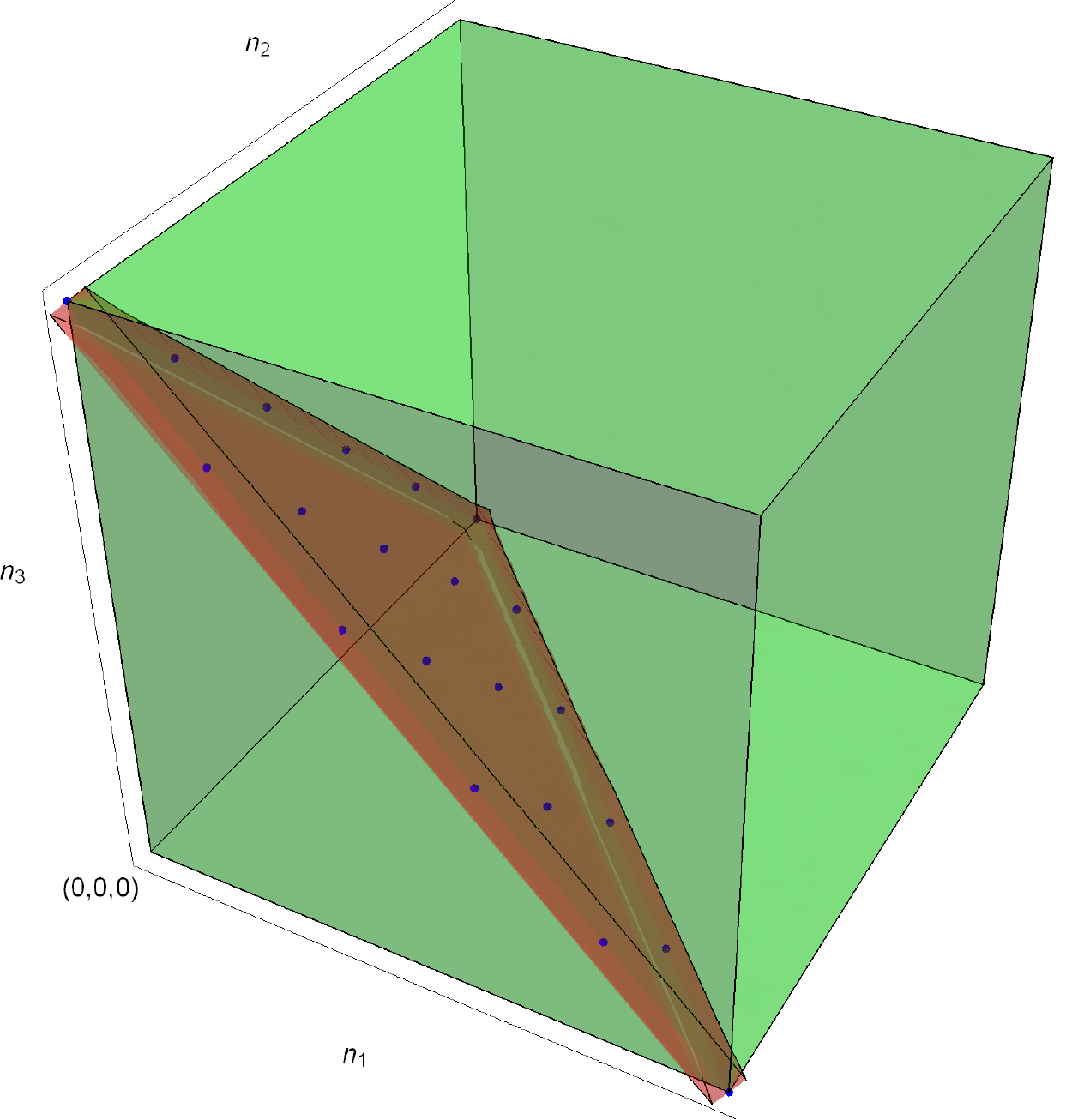} }}%
    \qquad
    \subfloat[\centering Exotic plot for $t=2,-2$]{{\includegraphics[width=6cm]{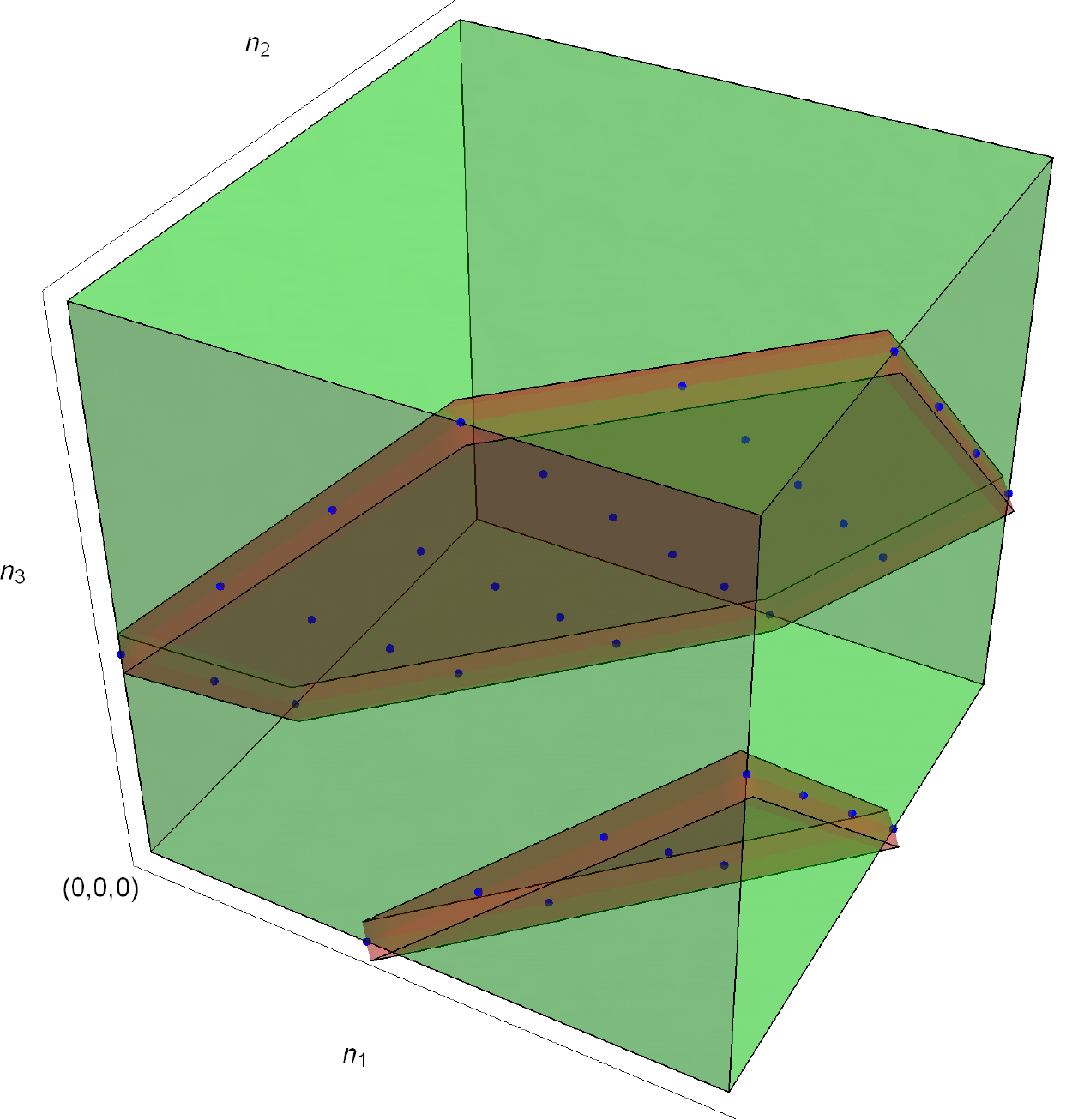} }}%
    \caption{We plot slices at different values of $t=\alpha(\mathfrak{m})$ of the octant of $\mathbb{R}^3$ spanned by positive $(n_1,n_2,n_3)$ such that $n_1=\mp(n_2+n_3-t)$. The orientation of the plane is related to the choice of fiber used to reduce. Each slice determines the eigenvalues contributing at the corresponding flux sector of the perturbative partition function on $\mathbb{CP}^2$.}%
    \label{plot:3dslices}%
\end{figure}

Considering the perturbative part at large $p$, we notice as before that the term $\mbox{mod }p$ in~\eqref{eq:projection.5d} becomes irrelevant. We are then free to set $t_{top}=\alpha(\mathfrak{m})$ and $t_{ex}=\alpha(\mathfrak{m})$. Equivalently we can impose:
\begin{equation}\begin{split}
    \mbox{top: }&\quad n_1=-n_2-n_3+\alpha(\mathfrak{m}),\\
    \mbox{ex: }&\quad n_1=+n_2+n_3-\alpha(\mathfrak{m}).
\end{split}\end{equation}
As in the previous section for $\mathbb{CP}^1$, the plots in \autoref{plot:3dslices} show the modes entering the perturbative partition function, at given $t$, on $\mathbb{CP}^2$. These correspond to slices of cones which are finite for the topologically twisted theory. Instead for the exotic theory one has to extended these slices to all positive $(n_1,n_2,n_3)$.\\ 

We can write the partition function on $\mathbb{CP}^2$ as:
\begin{equation}
    Z_{\mathbb{CP}^2}=\sum_{[\mathfrak{m}]}\int d\sigma_0 e^{-S_{cl}}Z^{pert}_{\mathbb{CP}^2}(\sigma_0,\mathfrak{m})Z^{non-pert}_{\mathbb{CP}^2}(\sigma_0,\mathfrak{m})
\end{equation}
Focusing on the perturbative partition function on $\mathbb{CP}^2$, we reduce the unfactorized expressions~\eqref{eq:unfactorized.5d.asd} along $x^{top}$ and~\eqref{eq:unfactorized.5d.flip} along $x^{ex}$:
\begin{equation}\begin{split}\label{eq:unfactorized.CP2.asd}
     Z_{\mathbb{CP}^2}^{pert,top}(\sigma_0,\mathfrak{m})=&\prod_{\alpha(\mathfrak{m})\geq n_2+n_3}\prod_{n_2,n_3\geq 0}\bigg(\epsilon_1 n_2+\epsilon_2 n_3+i\alpha(\sigma_0)+\bigg(1-\frac{\epsilon_1+\epsilon_2}{3}\bigg)\alpha(\mathfrak{m})\bigg)\\
     \times&\prod_{\alpha(\mathfrak{m})\geq n_2+n_3+1}\prod_{n_2,n_3\geq 1}\bigg(\epsilon_1 n_2+\epsilon_2 n_3+i\alpha(\sigma_0)+\bigg(1-\frac{\epsilon_1+\epsilon_2}{3}\bigg)\alpha(\mathfrak{m})\bigg).
\end{split}\end{equation}
\begin{equation}\begin{split}\label{eq:unfactorized.CP2.flip}
     Z_{\mathbb{CP}^2}^{pert,ex}(\sigma_0,\mathfrak{m})=&\prod_{\alpha(\mathfrak{m})\leq n_2+n_3}\prod_{n_2,n_3\geq 0}\bigg(\epsilon_1 n_2+\epsilon_2 n_3+i\alpha(\sigma_0)+\bigg(\frac{1}{3}-\frac{\epsilon_1+\epsilon_2}{3}\bigg)\alpha(\mathfrak{m})\bigg)\\
     \times&\prod_{\alpha(\mathfrak{m})\leq n_2+n_3+1}\prod_{n_2,n_3\geq 1}\bigg(\epsilon_1 n_2+\epsilon_2 n_3+i\alpha(\sigma_0)+\bigg(\frac{1}{3}-\frac{\epsilon_1+\epsilon_2}{3}\bigg)\alpha(\mathfrak{m})\bigg).
\end{split}\end{equation}
At the zero flux sector $\alpha(\mathfrak{m})=0$ the result for $Z_{\mathbb{CP}^2}^{pert,ex}(\sigma_0,\mathfrak{m})$ agrees with equation (108) in~\cite{Festuccia:2018rew}. Moreover, in both cases, we find for generic $\alpha(\mathfrak{m})$ a slice of cone, together with its interior, whose shape depends on the flux sector. These slices are exactly those pictured in \autoref{plot:3dslices} at different values of $t=\alpha(\mathfrak{m})$. We see how the modes contributing to each flux sector are finite for topologically twisted theories while infinite for exotic ones. This is consistent with the two cases being associated, respectively, to elliptic and transversely elliptic problems \cite{Festuccia:2018rew,Festuccia:2019akm}.  \\

Similarly we can reduce the factorized expression \eqref{eq:2.62asd} and \eqref{eq:2.62flip} to:
\begin{equation}\begin{split}\label{eq:CP2.factorized.asd}
    Z&_{\mathbb{CP}^2}^{pert,top}(\sigma_0,\mathfrak{m})=\prod_{i,j=0}^{\infty}\left(\omega_1\alpha(\mathfrak{m})+\epsilon_1 i +\epsilon_2 j+i\alpha(\sigma_0)\right)^{\frac{1}{2}}\left(\omega_1\alpha(\mathfrak{m})+\epsilon_1 (i+1) +\epsilon_2 (j+1)+i\alpha(\sigma_0)\right)^{\frac{1}{2}}\\
    &\left(\omega_2\alpha(\mathfrak{m})-\epsilon_1(-i-1)+(\epsilon_2-\epsilon_1) (-j-1)+i\alpha(\sigma_0)\right)^{\frac{1}{2}}\left(\omega_2 \alpha(\mathfrak{m})-\epsilon_1 (-i) +(\epsilon_2-\epsilon_1)(-j)+i\alpha(\sigma_0)\right)^{\frac{1}{2}}\\
    &\left(\omega_3 \alpha(\mathfrak{m})+(\epsilon_1-\epsilon_2)i-\epsilon_2 (-j-1)+i\alpha(\sigma_0)\right)^{-\frac{1}{2}}\left(\omega_3 \alpha(\mathfrak{m})+(\epsilon_1-\epsilon_2)(i+1) -\epsilon_2(-j)+i\alpha(\sigma_0)\right)^{-\frac{1}{2}}.
\end{split}\end{equation}
\begin{equation}\begin{split}\label{eq:CP2.factorized.flip}
   Z&_{\mathbb{CP}^2}^{pert,ex}(\sigma_0,\mathfrak{m})=\prod_{i,j=0}^{\infty}\left(-\omega_1 \alpha(\mathfrak{m})+\epsilon_1 i +\epsilon_2 j+i\alpha(\sigma_0)\right)^{\frac{1}{2}}\left(-\omega_1 \alpha(\mathfrak{m})+\epsilon_1 (i+1) +\epsilon_2 (j+1)+i\alpha(\sigma_0)\right)^{\frac{1}{2}}\\
    &\left(\omega_2 \alpha(\mathfrak{m})+\epsilon_1(-i-1)+(\epsilon_2-\epsilon_1) (-j-1)+i\alpha(\sigma_0)\right)^{\frac{1}{2}}\left(\omega_2 \alpha(\mathfrak{m})+\epsilon_1 (-i)+(\epsilon_2-\epsilon_1)(-j)+i\alpha(\sigma_0)\right)^{\frac{1}{2}}\\
    &\left(\omega_3 \alpha(\mathfrak{m})+(\epsilon_1-\epsilon_2)i+\epsilon_2 (-j-1)+i\alpha(\sigma_0)\right)^{-\frac{1}{2}}\left(\omega_3 \alpha(\mathfrak{m})+(\epsilon_1-\epsilon_2)(i+1) +\epsilon_2(-j)+i\alpha(\sigma_0)\right)^{-\frac{1}{2}}.
\end{split}\end{equation}
Our factorized results show that fluxes enter in the perturbative partition function simply as a shift in the Coulomb branch parameter. For topologically twisted and exotic theories we find the following shifts at each fixed point:
\begin{equation}\begin{split}
    &l=1\quad\mbox{top: }i\alpha(\sigma_0)+ \alpha(\mathfrak{m})\left(1-\frac{\epsilon_1+\epsilon_2}{3}\right)\;\;\quad \mbox{ex: }i\alpha(\sigma_0)+ \alpha(\mathfrak{m})\left(\frac{1}{3}-\frac{\epsilon_1+\epsilon_2}{3}\right),\\
    &l=2\quad\mbox{top: }i\alpha(\sigma_0)+ \alpha(\mathfrak{m})\left(1+\frac{2\epsilon_1-\epsilon_2}{3}\right)\quad \mbox{ex: }i\alpha(\sigma_0)+ \alpha(\mathfrak{m})\left(\frac{1}{3}+\frac{2\epsilon_1-\epsilon_2}{3}\right),\\
    &l=3\quad\mbox{top: }i\alpha(\sigma_0)+ \alpha(\mathfrak{m})\left(1+\frac{2\epsilon_2-\epsilon_1}{3}\right)\quad \mbox{ex: }i\alpha(\sigma_0)+ \alpha(\mathfrak{m})\left(\frac{1}{3}+\frac{2\epsilon_2-\epsilon_1}{3}\right).
\end{split}\end{equation}
Equations \eqref{eq:CP2.factorized.asd} and \eqref{eq:CP2.factorized.flip} are the main results of this paper as we are able to derive the factorized perturbative partition function on $\mathbb{CP}^2$ at all flux sectors for both $\mathcal{N}=2$ topologically twisted SYM and Pestun-like theories. This confirms the conjecture of \cite{Festuccia:2018rew} regarding the $\mathfrak{m}$-dependence at non trivial flux sectors. A more careful analysis would require also the study of instantons, which we hope to address in future work.\\

So far we have expressed the perturbative partition function as a super-determinant originating from the Gaussian integral around the localization locus:
\begin{equation}\label{eq:det.generic}
    \mbox{det }=\prod_i w_i^{-b_i},
\end{equation}
where $i$ captures all the eigenvalues under the transformations generated by the square of the fermionic generator. The corresponding eigenvalue is $w_i$ and the integers $b_i$ count the degeneracies in the modes at fixed eigenvalue. To connect with the results in~\cite{Festuccia:2018rew} we take the equivalent approach of computing the equivariant indices associated to the topologically twisted and exotic complexes\footnote{We refer to~\cite{Festuccia:2018rew,Festuccia:2019akm} for details on the topic.}. In general the index takes the form:
\begin{equation}\label{eq:index.generic}
    \mbox{ind }=\sum_i b_i e^{-w_i}.
\end{equation}
Hence, with some computations, we can use the relation between~\eqref{eq:det.generic} and~\eqref{eq:index.generic} to rewrite~\eqref{eq:CP2.factorized.asd} and~\eqref{eq:CP2.factorized.flip} as
\begin{equation}
    \mbox{ind }=-e^{i\alpha(\mathfrak{m})}\frac{I^+_{\alpha(\mathfrak{m})}+I^-_{\alpha(\mathfrak{m})}}{2}\chi_{adj}(e^{i\alpha({\sigma})}),
\end{equation}
where:
\begin{equation}\begin{split}\label{eq:index.+}
    &\mbox{top: }I^+_{\alpha(\mathfrak{m})}=\frac{e^{i\alpha(\mathfrak{m})(-\frac{\epsilon_1+\epsilon_2}{3})}}{(1-e^{i\epsilon_1})(1-e^{i\epsilon_2})}+\frac{e^{i\alpha(\mathfrak{m})(\frac{2\epsilon_1-\epsilon_2}{3})}}{(1-e^{-i\epsilon_1})(1-e^{i(\epsilon_2-\epsilon_1)})}+\frac{e^{i\alpha(\mathfrak{m})(\frac{2\epsilon_2-\epsilon_1}{3})}}{(1-e^{i(\epsilon_1-\epsilon_2)})(1-e^{-i\epsilon_2})},\\
    &\mbox{ex: }I^+_{\alpha(\mathfrak{m})}=\frac{e^{i\alpha(\mathfrak{m})(-\frac{\epsilon_1+\epsilon_2}{3})}}{(1-e^{i\epsilon_1})(1-e^{i\epsilon_2})}+\frac{e^{i\alpha(\mathfrak{m})(\frac{2\epsilon_1-\epsilon_2}{3})}}{(1-e^{i\epsilon_1})(1-e^{i(\epsilon_2-\epsilon_1)})}+\frac{e^{i\alpha(\mathfrak{m})(\frac{2\epsilon_2-\epsilon_1}{3})}}{(1-e^{i(\epsilon_1-\epsilon_2)})(1-e^{i\epsilon_2})},
\end{split}\end{equation}
\begin{equation}\begin{split}\label{eq:index.-}
    \mbox{top: }I^-_{\alpha(\mathfrak{m})}=&\frac{e^{i(\alpha(\mathfrak{m})-3)(-\frac{\epsilon_1+\epsilon_2}{3})}}{(1-e^{i\epsilon_1})(1-e^{i\epsilon_2})}+\frac{e^{i(\alpha(\mathfrak{m})-3)(\frac{2\epsilon_1-\epsilon_2}{3})}}{(1-e^{-i\epsilon_1})(1-e^{i(\epsilon_2-\epsilon_1)})}+\frac{e^{i(\alpha(\mathfrak{m})-3)(\frac{2\epsilon_2-\epsilon_1}{3})}}{(1-e^{i(\epsilon_1-\epsilon_2)})(1-e^{-i\epsilon_2})}\\
    =&I^+_{-\alpha(\mathfrak{m})}(-\epsilon_1,-\epsilon_2),\\
    \mbox{ex: }I^-_{\alpha(\mathfrak{m})}=&\frac{e^{i(\alpha(\mathfrak{m})-3))(-\frac{\epsilon_1+\epsilon_2}{3})}}{(1-e^{i\epsilon_1})(1-e^{i\epsilon_2})}+\frac{e^{i\alpha(\mathfrak{m})(\frac{2\epsilon_1-\epsilon_2}{3})}e^{i\epsilon_2}}{(1-e^{i\epsilon_1})(1-e^{i(\epsilon_2-\epsilon_1)})}+\frac{e^{i\alpha(\mathfrak{m})(\frac{2\epsilon_2-\epsilon_1}{3})}e^{i\epsilon_1}}{(1-e^{i(\epsilon_1-\epsilon_2)})(1-e^{i\epsilon_2})}\\
    =&I^+_{-\alpha(\mathfrak{m})}(-\epsilon_1,-\epsilon_2).
\end{split}\end{equation}
An important step in the previous computation is the regularization of $I^+_{\alpha(\mathfrak{m})}$ for topologically twisted theories and exotic theories (respectively $\mp$):
\begin{equation}\begin{split}\label{eq:regularization.index}
    &\left[\frac{1}{1-e^{i\epsilon_1}}\right]^+\left[\frac{1}{1-e^{i\epsilon_2}}\right]^+=\frac{1}{(1-e^{i\epsilon_1})(1-e^{i\epsilon_2})}=\sum_{j,k=0}^{\infty}e^{i(j\epsilon_1+k\epsilon_2)},\\
    &\left[\frac{1}{1-e^{\mp i\epsilon_1}}\right]^-\left[\frac{1}{1-e^{i(\epsilon_2-\epsilon_1)}}\right]^-=\frac{1}{(1-e^{\mp i\epsilon_1})(1-e^{i(\epsilon_2-\epsilon_1)})}=\sum_{j,k=0}^{\infty}e^{-i((j+1)(\mp\epsilon_1)+(k+1)(\epsilon_2-\epsilon_1))},\\
    &\left[\frac{1}{1-e^{i(\epsilon_1-\epsilon_2)}}\right]^+\left[\frac{1}{1-e^{\mp i\epsilon_2}}\right]^-=\frac{1}{(1-e^{i(\epsilon_1-\epsilon_2)})(1-e^{\mp i\epsilon_2})}=-\sum_{j,k=0}^{\infty}e^{i(j(\epsilon_1-\epsilon_2)-(k+1)(\mp\epsilon_2))}.
\end{split}\end{equation}
We recall that the regularization for the topologically twisted and exotic theories is the same even if some local equivariant parameters do not come with the same sign. For both cases $I^-_{\alpha(\mathfrak{m})}(\epsilon_1,\epsilon_2)=I^+_{-\alpha(\mathfrak{m})}(-\epsilon_1,-\epsilon_2)$ and it is enough to switch regularization and $\epsilon_{1,2}$-dependence in~\eqref{eq:regularization.index}. Then as final result we find, at trivial flux sector, for topologically twisted and exotic theories:
\begin{equation}\begin{split}
    &\mbox{top: }\left[\frac{1}{1-e^{i\epsilon_1}}\right]^+\left[\frac{1}{1-e^{i\epsilon_2}}\right]^+ + \left[\frac{1}{1-e^{- i\epsilon_1}}\right]^-\left[\frac{1}{1-e^{i(\epsilon_2-\epsilon_1)}}\right]^- + \left[\frac{1}{1-e^{i(\epsilon_1-\epsilon_2)}}\right]^+\left[\frac{1}{1-e^{- i\epsilon_2}}\right]^- + c.c.\\
    &\mbox{ex: }\left[\frac{1}{1-e^{i\epsilon_1}}\right]^+\left[\frac{1}{1-e^{i\epsilon_2}}\right]^+ + \left[\frac{1}{1-e^{ i\epsilon_1}}\right]^-\left[\frac{1}{1-e^{i(\epsilon_2-\epsilon_1)}}\right]^- + \left[\frac{1}{1-e^{i(\epsilon_1-\epsilon_2)}}\right]^+\left[\frac{1}{1-e^{ i\epsilon_2}}\right]^- + c.c.
\end{split}\end{equation}
These equations match with the results in~\cite{Festuccia:2018rew} for topologically twisted and flip theories\footnote{In~\cite{Festuccia:2018rew} flip and flip' theories are related by $\epsilon_{1,2}\rightarrow-\epsilon_{1,2}$.} after identifying $e^{i\epsilon_1}=s$ and $e^{i\epsilon_2}=t$. Notice that the for the topologically twisted theory we are computing the index for an elliptic operator and consistently we find that all terms cancel except $2e^{(0\epsilon_1+0\epsilon_2)}=2$.

\section*{Acknowledgments}
We thank Guido Festuccia and Jian Qiu for illuminating discussions. We also thank Guido Festuccia, Joseph Minahan, Luigi Tizzano and Maxim Zabzine for critical readings of the draft. L.R. acknowledges the support of the ERC STG Grant 639220 and of Vetenskapsrådet under grant 2018-05572. 

\appendix

\section{Properties of multiple sine functions}\label{app:one}
We define:
\begin{equation}
	\boldsymbol{\omega}=(\omega_1,...,\omega_r)=(1+a_1,...,1+a_r),\quad \omega_1,...,\omega_r\in\mathbb{R}.
\end{equation}
We define the multiple zeta function as:
\begin{equation}
	\zeta_r(s,z|\boldsymbol{\omega})=\sum_{n_1,...,n_r=0}^{\infty}\frac{1}{(n_1\omega_1+...+n_r\omega_r+z)^s},
\end{equation}
where $z\in\mathbb{C}$ and $\mbox{Re }s>r$. The series can be analytically continued to the complex plane, due to holomorphicity in the domain. Thus we can introduce the multiple gamma function:
\begin{equation}
	\Gamma_r(z|\boldsymbol{\omega})=\exp\bigg(\frac{\partial}{\partial s}\zeta_r(s,z|\boldsymbol{\omega})\big|
_{s=0}\bigg),
\end{equation}  
as a building block of the multiple sine function:
\begin{equation}
	S_r(z|\boldsymbol{\omega})=\Gamma_r(z|\boldsymbol{\omega})^{-1}\Gamma_r(\omega_{tot}-z|\boldsymbol{\omega})^{(-1)^r}.
\end{equation}
The multiple sine function can alternatively be expressed in the following ways:
\begin{equation}\begin{split}\label{eq:Sr.factorization}
    S_r(z|\boldsymbol{\omega})=&e^{(-1)^r\frac{\pi i}{r!}B_{r,r}(z|\boldsymbol{\omega})}\prod_{i=1}^r\Big(e^{2\pi i \frac{z}{\omega_i}}; e^{2\pi i \frac{\omega_1}{\omega_i}},\ldots\lor_i\ldots,e^{2\pi i \frac{\omega_r}{\omega_i}}\Big)\\
    =&e^{(-1)^{r-1}\frac{\pi i}{r!}B_{r,r}(z|\boldsymbol{\omega})}\prod_{i=1}^r\Big(e^{-2\pi i \frac{z}{\omega_i}}; e^{-2\pi i \frac{\omega_1}{\omega_i}},\ldots\lor_i\ldots,e^{-2\pi i \frac{\omega_r}{\omega_i}}\Big).
\end{split}\end{equation}
Where the $B_{rr}$ is the Bernoulli polynomial, defined by the following generating function:
\begin{equation}
    \frac{t^re^{zt}}{\prod_{j=1}^r(e^{\omega_j}-1)}=\sum_{n=0}^\infty B_{r,n}(z|\boldsymbol{\omega})\frac{t^n}{n!}.
\end{equation}
We also carry the following definitions of the q-Pochhammer symbols:
\begin{equation}\label{eq:q.Pochhammer}
    (z;q_1,\ldots,q_{k},q_{k+1},\ldots,q_r)=\Bigg[\prod_{j_1,\ldots,j_r=0}^{\infty}\big(1-zq_1^{-j_1-1}\dots q_k^{-j_k-1}q_{k+1}^{j_{k+1}}\dots q_r^{j_r}\big)\Bigg]^{(-1)^k},\qquad \begin{matrix}|q_i|>1\quad 1\leq i\leq k\\
    |q_i|<1 \quad k<i\leq r
    \end{matrix}
\end{equation}
For proofs of above mentioned identities and conventions used we refer to \cite{narukawa2004modular}.\\

Let us consider the perturbative part of the partition function of a vector multiplet coupled to matter in a representation $R$ of the gauge group $G$. For both $S^3$ and $S^5$ the one-loop determinant can be written in terms of multiple sine functions modulo $\sigma$-independent factors:
\begin{equation}
	Z^{2r-1}_{1-loop}=\frac{\prod_{\alpha\in roots} S_r(i\alpha(\sigma)|\boldsymbol{\omega})}{\prod_{\rho\in R} S_r(i\rho(\sigma)+\frac{\omega_{tot}}{2}|\boldsymbol{\omega})}.
\end{equation}

\bibliographystyle{utphys}
\bibliography{Moddbib}

\end{document}